\documentclass[]{JHEP}
\usepackage{epsfig}

\newcommand{\fourgraphs}[4]{%
\unitlength=1in
\begin{picture}(6,5)
\put(0,0){\epsfig{file=#3.eps, width=3in}}
\put(3,0){\epsfig{file=#4.eps, width=3in}}
\put(0,2.5){\epsfig{file=#1.eps, width=3in}}
\put(3,2.5){\epsfig{file=#2.eps, width=3in}}
\put(-0.1,2){(c)}
\put(-0.1,4.5){(a)}
\put(2.9,2){(d)}
\put(2.9,4.5){(b)}
\end{picture}}

\newcommand{\twographs}[2]{%
\unitlength=1in
\begin{picture}(6,2.5)
\put(0,0){\epsfig{file=#1.eps, width=3in}}
\put(3,0){\epsfig{file=#2.eps, width=3in}}
\put(-0.1,2){(a)}
\put(2.9,2){(b)}
\end{picture}}

\newcommand{\bmat}{\left(\begin{array}}
\newcommand{\emat}{\end{array}\right)}

\def\yzero{\smash{\hbox{$y\kern-4pt\raise1pt\hbox{${}^\circ$}$}}}

\def\beq{\begin{equation}}
\def\eeq{\end{equation}}
\def\beqa{\begin{eqnarray}}
\def\eeqa{\end{eqnarray}}

\def\-{\hphantom{-}}

\def\s2{\frac{1}{\sqrt2}}

\def\beq{\begin{equation}}
\def\eeq{\end{equation}}
\def\beqa{\begin{eqnarray}}
\def\eeqa{\end{eqnarray}}

\def\IF{\relax{\rm I\kern-.18em F}}
\def\II{\relax{\rm I\kern-.18em I}}
\def\IP{\relax{\rm I\kern-.18em P}}
\def\IC{\relax\hbox{\kern.25em$\inbar\kern-.3em{\rm C}$}}
\def\IR{\relax{\rm I\kern-.18em R}}

\def\cp{{\cal P}}

\def\Dsl{\,\raise.15ex\hbox{/}\mkern-13.5mu D} 
\def\IZ{Z\kern-.4em  Z}

\def\cp#1{\relax\ifmmode {\IP\kern-2pt{}_{#1}}\else
$\IP\kern-2pt{}_{#1}$\fi}

\newcommand{\cO}{{\cal O}}

\newcommand{\hf}{\frac12}

\newcommand{\bea}{\begin{eqnarray}}
\newcommand{\eea}{\end{eqnarray}}
\newcommand{\be}{\begin{equation}}
\newcommand{\ee}{\end{equation}}
\newcommand{\bt}{\begin{tabular}}
\newcommand{\et}{\end{tabular}}
\newcommand{\ba}{\begin{array}}
\newcommand{\ea}{\end{array}}

\newcommand{\R}{{\rm Re}}

\newcommand{\mgr}{m_{3/2}}
\newcommand{\agut}{\alpha_{\rm GUT}}
\newcommand{\leqsim}{\,\raisebox{-0.6ex}{$\buildrel < \over \sim$}\,}
\newcommand{\geqsim}{\,\raisebox{-0.6ex}{$\buildrel > \over \sim$}\,}

\def \half{{\textstyle\hf}}

\def \soll={\stackrel{!}{=}}
\def \rf=#1{\stackrel{(\ref{#1})}{=}}

\def \dgs{\delta_{\rm GS}}

\def \gev{{\mbox{~GeV}}}

\def \w{{\rm with\ }}

\def \eg{{\it e.g.\ }}
\def \ie{{\it i.e.\ }}

\preprint{DAMTP-2000-22}

\title{Soft SUSY Breaking, Dilaton Domination and Intermediate Scale String Models}

\author{S.A. Abel \\ LPT, Universit\'e Paris-Sud, Orsay,
91405, France}
\author{B.C. Allanach, F. Quevedo \\ DAMTP, Wilberforce Rd, Cambridge CB3 0WA, UK}
\author{L. Ib\'a\~nez, M. Klein \\ Departamento de F\'{\i}sica Te\'orica C-XI
and Instituto de F\'{\i}sica Te\'orica  C-XVI,
Universidad Aut\'onoma de Madrid,
Cantoblanco, 28049 Madrid, Spain.}


\abstract{
We present an analysis of the low-energy implications of 
an intermediate scale ($\sim 10^{11}$ GeV) string theory.
 We mainly focus on 
 the evolution of the physical parameters under the
renormalisation group equations (RGEs) and 
find  several  interesting new features that differ from the
standard GUT scale or Planck scale scenarios.
We give a general discussion of soft supersymmetry 
breaking terms in type I
theories and then investigate the renormalization group running.
In the dilaton domination scenario, we
present the sparticle spectra,
analysing constraints from charge and colour breaking, fine tuning and
radiative electroweak symmetry breaking. 
We compare with the allowed regions of parameter space when the RGEs start
running at the standard GUT or the intermediate scales, and find
quite remarkably that the dilaton dominated supersymmetry breaking scenario, 
which is essentially ruled out from
constraints on charge and colour breaking if the fundamental scale is
close to the Planck mass, is allowed in a large region of
parameter space if the fundamental scale is intermediate. 
}

\begin{document}
\section{Introduction}

Recently we have seen a radical change in our
understanding of the possible
physical implications of a fundamental theory.
If such a theory includes higher dimensional objects, such as 
D-branes, then the fundamental scale of the theory can be very 
small compared to the Planck scale
\cite{witten,lykken,add,aadd}. This is because
if some or all of the observable matter and interactions is confined to
a brane (our universe) which is embedded in a higher dimensional bulk
world, then the fundamental scale becomes essentially a free parameter.

One could say that part of the recent progress in string theory is
the realization  that the fundamental scale of the theory is actually 
completely unknown, whereas it was previously thought that it had to be 
near the Planck scale if strings were to describe gravity. This new 
freedom arises mainly because the size of the extra
dimensions is not fixed within the theory, a manifestation of
the vacuum degeneracy problem in conjunction with the world as a 
brane scenario.

This ignorance has been turned into a virtue since now we may use
physical arguments to motivate possible values of the
fundamental scale. For instance we may entertain
the idea that the fundamental scale can be as low as the current
experimental limits allow, {\it ie}\/ 1 TeV
\cite{lykken,add,aadd}. Two other, less radical
proposals have been put forward as well. The first assumes that the size of
the extra dimensions is such that the fundamental scale coincides with
the GUT scale \cite{witten},
 automatically solving the problem of gauge coupling
unification in string theory, which is so far the main indirect
experimental information we have about  physics at higher energies.
The second proposal sets the string scale to the intermediate value
$M_s\sim 10^{10-12}$ GeV \cite{benakli,biq}. This choice is 
motivated by several indicators, most
notably the scale of supersymmetry breaking in hidden sector or
gravity mediated scenarios.

Lowering the string scale may have spectacular  implications at low
energies, many of which are currently being explored in great detail.
The TeV scale scenario is clearly the most studied since it is closer
to the experimental limits and has direct implications for LHC
physics. Here we will consider in detail the
intermediate scale scenario, and will show that it too 
can have important low-energy
implications. First we will review the motivations for
introducing such a scale, then we will study its phenomenological
implications mainly in the context of type I string models
although similar results may be obtained e.g.\ in the type IIB 
non-orientifold orbifold models  recently constructed in 
\cite{aiqu}.

The most important issue that we will address is the running of the 
physical parameters with the renormalization group. In the standard
approach, all of the physically relevant parameters such as the 
gauge couplings and the soft supersymmetry breaking terms start
running from a very high scale, namely the GUT or Planck scale. An
enormous amount of knowledge has been accumulated about the various
implications of this running, particularly the constraints on the
values of the soft supersymmetry breaking parameters. Our main goal in
this article is to begin to see how this analysis changes when we
start the running from the intermediate scale. 
This simple modification may have very interesting
implications for the unification of gauge
couplings, $b-\tau$ unification, the possible quasi-fixed points, 
and so on.
We will focus mainly on the running of the soft supersymmetry breaking
terms in order to study a wide set of standard constraints coming from
fine tuning of parameters thus achieving electroweak  symmetry breaking, 
imposing the absence of electric charge and $SU(3)$
colour breaking etc.

Our work ought to be considered as only a starting point on
the phenomenological issues of the intermediate scale scenario
motivated by strings. Recently there has been some progress in the 
construction of phenomenologically realistic brane models with
supersymmetry explicitly broken \cite{aiq}.
 The concrete models differ in several
ways from the standard MSSM scenarios. Besides having unification at
the intermediate scale, the hypercharge normalization is generically
also different from the standard $5/3$ GUT inspired value. In the
present article we will not consider explicit models but rather 
will try to analyze the intermediate scale scenario in general. 
The only departure we make from the MSSM structure is that
we assume gauge coupling unification takes place 
at the fundamental, intermediate, scale. To achieve this unification we
will consider the simplest possibility of adding several lepton pairs to
accelerate the $SU(2)\times U(1)$ running and cause precocious
unification \cite{biq}.
 Other proposals \cite{keith}, such as mirage unification
\cite{mirage},
 may achieve
this unification without requiring extra matter fields.
We will examine both cases and show that for most phenomenological
purposes there is little difference between them.
We leave for a future publication the consideration of further
departures from the MSSM such as the changing of $U(1)$ normalization
as suggested in ref.\cite{aiq}.

In the next section we review the arguments in favour of an
intermediate
scale. Then in section 3 we discuss general issues concerning 
the structure of type I models. 
In section 4 we discuss soft supersymmetry breaking terms mainly
in the context of type I strings, which allow the possibility of
lowering the string scale. Assuming that SUSY-breaking is transmitted
by the closed string sector of the theory, 
 we find general expressions for the soft
breaking parameters in terms of the $F$ terms of the moduli fields,
the dilaton $S$, compactification size moduli $T_i$ and twisted
blowing-up fields $M$. In section 5 we discuss the implications
of these modifications for the physically relevant questions mentioned above.

\section{The Case for the Intermediate Scale Scenario}

\subsection{Supersymmetry Breaking}

The origin of the intermediate scale may be traced back to the
early 1980's when studies of supersymmetric models showed that
the most efficient way to break supersymmetry was in a hidden sector.
In those days, the preferred and simplest transmission of 
the information of supersymmetry
breaking to the observable world was via gravitational interactions,
giving rise to the `hidden sector' or `gravity mediated' scenarios.
In these scenarios, 
because the observable sector only knows about supersymmetry breaking
through
gravitational strength interactions, the splitting among 
supersymmetric multiplets is of order 
$M_{SUSY}^2/M_{Planck}$. If supersymmetry is to solve the
hierarchy problem this splitting should be close to the electroweak
scale thereby fixing the scale of supersymmetry breaking to be 
of order  
\beq
M_{SUSY}\sim \sqrt{M_W\ M_{Planck}}\sim 10^{10-12} \gev.
\eeq
 This is the most
studied supersymmetry breaking scenario of the past 18 years.
An alternative  is the `gauge mediated' scenario in which
gauge interactions rather than gravity connect the hidden and
observable sectors and the supersymmetry breaking scale is
therefore 
close to the electroweak scale. Other possibilities have been introduced
more recently, a particularly interesting one being the `anomaly
mediation' scenario.
As is the case for the gravity mediated
supersymmetry breaking, these contributions to the supersymmetry
breaking in the observable sector are always present.

In perturbative heterotic strings the hidden sector scenario could be
nicely realized since the gauge group was $E_8\times E_8$.
The standard model particles could be charged under only
one of the two $E_8$ groups with the supersymmetry breaking hidden
sector being charged under the second $E_8$. Gravity would then be the
messenger of supersymmetry breaking to the observable sector.
However, in this framework the scale of
string theory was close to the Planck scale and the intermediate scale
appeared only as a low energy phenomenon, being the scale at which the
hidden gauge interactions become strong. In this class of models
supersymmetry
was broken via some non-perturbative field theoretical effect such as
gaugino condensation, whereas the string theory was only treated at
the perturbative level due to the lack of understanding of
non-perturbative string effects.

Non-perturbative string effects are now beginning to be understood, thanks 
mainly to the discovery of D-branes (surfaces on which the endpoints of
open strings are attached) in type I and type II strings, as well
as the Horava-Witten formulation of the heterotic string which starts
from 11 dimensional 
supergravity compactified on the interval $S_1/\IZ_2$, with each
of the $E_8$'s living at the endpoints of the interval. 

D-branes participate in supersymmetry breaking. First, being BPS
configurations, they partially break supersymmetry. Second, in the
compactification process they may wrap around different topologically
non-trivial cycles of the compact space. Depending on the nature of
these cycles, supersymmetry may or may not be further
broken. Furthermore,
there has been recent activity on non-BPS brane configurations that
tend to break all supersymmetries. We can therefore imagine a situation
with many different brane configurations, some of them breaking
supersymmetry partially or completely, with a generic vacuum being
non-supersymmetric. It is then natural for the fundamental string
scale to be either the intermediate scale, if we live on  a
supersymmetric brane and supersymmetry is only broken by 
other distant branes, or the TeV scale if supersymmetry is
broken in our brane. 
We naturally expect the fundamental scale 
to be of the same order or smaller than the intermediate scale since,
if it were much larger, we would feel the supersymmetry breaking 
too strongly and would be faced with 
the hierarchy problem. 
In this case the supersymmetry breaking of the string physics
can also play the role of low-energy supersymmetry breaking.

The brane/anti-brane models constructed recently realize these
possibilities. The TeV
scale scenario can be constructed with supersymmetry being either broken 
explicitly on our brane or communicated
by gauge mediation. However, even though there are suggestions for
addressing issues such as proton stability and gauge unification
\cite{keith}
in this scheme, so far there are no convincing
concrete models for the TeV scenario.
On the other hand it is much simpler to realize 
the intermediate scale scenario whilst satisfying 
both of these requirements since, for instance,
unification can still be achieved logarithmically, and generically
baryon and lepton number violating operators 
are much more suppressed.

\subsection{Strong CP Problem}

The symmetries of the Standard Model allow for a term
\beq
{\cal{L}}
_{\bar{\Theta}}\ =\ \frac{g^2}{32\pi^2}\ \bar{\Theta }\,
F^{\mu\nu} 
\tilde{F}_{\mu\nu},
\eeq
where $F^{\mu\nu}$ is the QCD field strength and $\bar{\Theta}$ is an
arbitrary parameter that essentially reflects the non-trivial nature
of the QCD vacua 
and the fact that the Standard Model is chiral. 
This term explicitly violates CP and 
consequently $\bar{\Theta}$ is strongly bounded 
by the experimental limits on the neutron electric dipole moment,
\beq
\bar{\Theta}<10^{-9}.
\eeq 
This unnaturally small bound is the strong CP problem. 
The most elegant solution to it is the Peccei-Quinn mechanism in
which an additional chiral symmetry, $U(1)_{PQ}$, is added to 
the model and broken spontaneously. The corresponding Goldstone
mode of this symmetry is the axion field and the 
static $\bar{\Theta}$ parameter is substituted by a dynamical
one, $a(x)/f_a$, where $a(x)$ is the axion field
and $f_a$ is a dimensionful constant known as the axion decay
constant. The non-perturbative dynamics of QCD
then fixes the axion field to be at the minimum of its
potential, $a=0$, thereby solving the strong CP problem
(for a review see \cite{kim}).

The axion decay constant, $f_a$ is very strongly constrained, mostly by
astrophysical and cosmological bounds. Requiring that 
axion emission does not over-cool stars gives lower 
bounds on $f_a$. These have
been determined for many different systems, 
from the sun to red giants and
globular clusters. The
strongest constraint comes from supernova SN1987a which requires
$f_a>10^9 \gev$ \cite{astroaxion}. 
Since the axion couples so weakly to normal matter its lifetime
exceeds the age of the universe by many orders of magnitude, 
and it also has a low  mass
($m_a\sim \Lambda_{QCD}^2/f_a$). 
Consequently the axion also has interesting cosmological
implications, especially as a cold dark matter candidate. 
Indeed coherent oscillations around the minimum of its potential may dominate
the energy density of the universe if its potential is very
flat. This puts a lower bound on the axion mass which
leads to an upper bound for $f_a$ of order
$f_a \leqsim 10^{12} \gev$. Combining the astrophysical and cosmological bounds
gives a very narrow window for the axion decay constant;
\beq
10^9\ \gev\ \leqsim \ f_a\ \ \leqsim\ 10^{12}\ \gev.
\eeq

Therefore if we want the Peccei-Quinn mechanism to work, we must
explain the value of $f_a$. The allowed range is remarkably consistent
with the intermediate scale, thus providing a strong argument 
in favour of the   intermediate string scale scenario.
Alternative scenarios generally have great difficulty in explaining 
why $f_a$ falls precisely within its narrow allowed window.
We note that there exist invisible axion models~\cite{axionX} for the TeV
scale extra dimension scenario. In these models, the mass of the axion is
independent from the scale of Peccei-Quinn symmetry breaking~\cite{axionX}.

In the mid 1980's, one of the most interesting phenomenological
arguments in favour of string theory was precisely the fact that these
theories always 
predicted axion fields. There are two 
kinds of string
axions in perturbative heterotic strings. The first is the 
model independent axion
coming from the imaginary component of the complex dilaton field, $S$
($a= {\rm{Im}} S$),
which is always present in string theory. The second kind of axion 
is model dependent and is associated with the moduli $T$ fields, which 
measure the
size and shape of the extra 6D compact space. These clearly depend on
the compactification. However none of these fields seems to be the
required QCD axion. The main obstacle for the perturbative 
heterotic string is the value of the axion decay
constant, which is 
constrained to be close to the Planck scale. 
Moreover, the model dependent axions do
not have the right couplings to start with and their corresponding
Peccei-Quinn symmetry is not preserved by world-sheet
instanton  corrections.
On the other hand  the model 
independent axion couples not only to QCD but also to the
hidden sector gauge fields in a universal way;
\beq
{\cal {L}}_{axion}\ = \frac{a}{M_P}\left( \left[ F^{\mu\nu} \tilde
F_{\mu\nu}\right]_{QCD}+\left[ F^{\mu\nu} \tilde
F_{\mu\nu}\right]_{hidden}\right).
\eeq
The non perturbative dynamics of the hidden sector is
then almost certainly the main source of the axion potential 
and the QCD contributions are essentially irrelevant.
Therefore none of the axions present in the heterotic string 
is able to solve the strong CP problem of QCD.

Recent studies of, for example, type I strings have changed the above
picture in a radical way. First, as we mentioned above, the string scale does
not have to be similar to the Planck scale but can be as low as we
want.
Therefore type I realizations of the intermediate scale scenario have
the right axion decay constant to satisfy the astrophysical and
cosmological requirements. Furthermore, in these models there are many
new candidates for axion fields which also couple to $F^{\mu\nu} \tilde
F_{\mu\nu}$. These fields are Ramond-Ramond fields $M_\alpha$
 associated with the
blowing-up of orbifold singularities in orientifold constructions.
Since the complex gauge coupling 
takes the form $f_i=S+s_i^\alpha M_\alpha$, where $i$
labels the different gauge groups,
{\em different}\/ combinations of these fields with the model independent
axion couple to QCD and the hidden sector groups.
This evades the second problem of the string axions mentioned above.

\subsection{Other Arguments}

In refs.\cite{benakli,biq} several other arguments were given in
favour of the intermediate scale scenario. First the realization of
the see-saw mechanism for neutrino masses is consistent with a
fundamental intermediate scale. 
It is also possible to accommodate neutrino masses in
TeV scale extra dimensions~\cite{neutX} via power law Yukawa coupling
suppression or Kaluza-Klein see-saw.
Cosmologically, several models of
chaotic inflation prefer the intermediate scale~\cite{lindekaloper}.
Finally the observed
ultra-high energy cosmic rays, which have energies 
of order $10^{20}$~eV, could be the products of 
string mode decays
if the fundamental scale is intermediate. These string modes are also 
good candidates for non-thermal dark matter known as 
wimpzillas \cite{wimpzillas}.

Concerning 
the arguments {\em against}\/ any scale below $M_{GUT}$
the two leading arguments against are firstly that the
only experimental indication we have for higher energies is the
apparent
joining of the strong and electroweak coupling constants at $M_{GUT}$
in the MSSM, and secondly that it is
difficult to obtain proton stability with models of lower fundamental
scales\footnote{Notice in this respect that in string models with 
a large string scale $M_s\propto M_X =2\times 10^{16}$ GeV, 
proton stability is in general also a problem. Even if one manages to 
get a model with R-parity which forbids lepton and/or 
baryon number violating d=4 operators, generic dim=5 operators
yield still too much proton decay unless they are forbidden by additional
symmetries.}.
However, as mentioned in the introduction, explicit type I string
models have recently been constructed where unification and proton
stability are achieved  with an intermediate fundamental 
scale \cite{aiq}. 

All the arguments given in this section make intermediate 
models serious alternatives to 
the MSSM GUT scale unification scenario, and in later sections
we shall see that they have other  
phenomenological benefits as well. 

\section{Structure of Type I String Models}

Most discussions of string phenomenology in the past 
focused on the heterotic string, since this model appeared
to have the best phenomenological properties. However following the
discovery of D-branes and string dualities, we have come to appreciate
the richness and phenomenological qualities of type I models. Let us
briefly discuss their main features of relevance for the rest of the
paper.

Type I string models can be constructed by starting with the
type IIB theory and performing a so-called orientifold twist      
$\Omega$ corresponding to the $Z_2$ identification of 
the two different orientations of the closed type IIB string
\cite{orient}.
Open strings appear as kind of twisted sectors under this operation\footnote{The form of the orientifold operation is 
related to the type of Dp-branes in the model. Thus e.g.\
for a $Z_N$  orientifold with D3-branes, it is
$\Omega (-1)^{F_L}R_1R_2R_3$, instead of just $\Omega $ 
for D9-branes. Here $R_i$ are reflection operators with respect to the
three complex compact planes and $F_L$ is the left-handed fermion number
of the Type IIB string.}
and are required in order to cancel all tadpoles of the twisted
theory. Compactification to four dimensions can then be 
achieved by a standard orbifold
twist in the six extra dimensions  giving rise to models with $N=1$
supersymmetry in four dimensions.
Solitonic objects of the type IIB theory correspond to extended
objects
known as D-branes where the endpoints of the open strings are
attached.
In order to preserve supersymmetry there can be only D-branes of
dimensions differing by a multiple of 4. So a generic compactification
has for instance 3 and 7-branes, with different gauge groups on each.
$T$-duality with respect to the 3 complex extra dimensions
exchanges D3 branes with D9-branes and D7-branes with  D5-branes.
One can accommodate e.g., the standard model group inside 
D3-branes and different quarks and leptons will come from the
exchange of open strings in between D3-branes  or between
D7-branes and D3-branes.

The $N=0$ models have a similar construction but include 
the additional feature of anti-branes which break supersymmetry,
with some
of the branes and anti-branes being required to live at the orbifold
fixed points in order to cancel tadpoles
\cite{ads}. In these models
supersymmetry breaking in the anti-branes is transmitted to the
observable branes via gravitational interactions (for which we 
require that the fundamental scale be the intermediate scale
as explained in the previous section) 
or directly for which the fundamental scale has to be
close to the electroweak symmetry breaking energy scale.

In these models, as in the heterotic models, there is always a dilaton
field and an antisymmetric tensor field, which in four-dimensions combine to
make a single chiral superfield $S$ (after appropriate dualization of
the antisymmetric tensor field to a scalar). There are also moduli
fields
$T$ associated with the size and shape of the extra six dimensions.
The explicit expression for these fields in terms of the string scale
$\alpha'$, the string coupling $\lambda=e^{-\phi}$ and the 
orbifold compactification radii
$R_i$ depends on the particular brane configuration.
For 3-branes, these fields can be written as \cite{imr}
\beqa
S & = & \frac{2}{\lambda }\ +\
i\theta\nonumber\\
T_i & = & \frac{2R_j^2R_k^2}{\lambda\alpha'^2}\ + \ i \eta_i,  \
\ i\not= j\not=k
\eeqa
where
 $\theta$ and $\eta_i$ are untwisted Ramond-Ramond closed string
states (dual to antisymmetric tensors).
For other brane configurations the expressions for these fields can
 easily be obtained by using $T$-duality for each of the three complex
 dimensions. These transformations can for instance switch the role of
$S$ and one of the $T_i$ fields, allowing for the possibility that 
$T_i$ rather than $S$ plays the role of the 
standard model gauge coupling constant.
This new degree of freedom opens up several interesting possibilities
 as discussed in ref.\cite{imr} of allowing different gauge groups to live
 in different branes and so have different gauge coupling functions.
 However here we will restrict the discussion to the cases where 
 all gauge fields live on a single
 brane and the effect of $T$ duality just amounts to a relabelling of the
 fields $S$ and $T_i$.

In orientifold models there are extra fields which also come from
Ramond-Ramond antisymmetric tensors and combine with the blowing-up
modes into full chiral multiplets. These fields, usually denoted as
$M_\alpha$
play important roles in cancelling $U(1)$ anomalies and generating
Fayet-Iliopoulos terms, in contrast with the heterotic case where only the
dilaton
plays that role. These are precisely the fields mentioned in the
previous section which
provide good stringy candidates for axion fields.
The holomorphic gauge function takes the general form, for a
$Z_N$
orientifold model at the disk level
\cite{iru,abd},
\beq
f_a\ = \ S\ + \ \sum_\alpha\, s_{a}^\alpha\ M_\alpha,
\eeq
where the $s_a^\alpha$ are computable 
model-dependent constants. The gauge coupling is
given by ${\rm Re} f_a=4\pi/g_a^2$.

Let us briefly see how in type I models the string scale can be as
small as is allowed by experiment.
For a configuration with the standard model spectrum belonging to 
a Dp-brane, the low energy action takes the form
\beq
S\ =\ -\frac{1}{2\pi} \int d^4x\sqrt{-g}\left(\frac{R^6\
M^8}{\lambda^2}{\cal R}\ +\ \frac{(RM)^{p-3}}{4\lambda}\ F_{\mu\nu}^2\
+\ \cdots \right),
\eeq
where $M=1/\sqrt{\alpha'}$ is the type I string scale, $R$ is taken as
the overall size of the compact 6D space and $\lambda$ is the dilaton.
Comparing the coefficient of the Einstein term with the physical
Planck mass $M_{Planck}^2$ and the coefficient of the gauge kinetic
term with the physical gauge coupling constant $\alpha_p$($\sim 1/24$
at the string scale), we find the relation
\beq
 M^{7-p}\ =\ \frac{\alpha_p}{\sqrt 2}\ M_{Planck}\ R^{p-6}.
\eeq
{}From this relation we can easily see that if the Standard Model
fits inside   
D3-branes  we may have $M\sim 10^{11}$ GeV
as long as the radius of the internal space is as large as 
  $R\sim 10^{-23}cm$
\cite{biq,imr}. Therefore the
required radii are large compared with the Planck length but still
extremely small compared with  $1mm$ as required in some cases with
$M\sim 1$ TeV. Notice that this analysis does not work for the
perturbative heterotic string since in that case the relation between 
$M_{Planck}$ and $M$ is independent of the size of the extra
dimensions
$M=\sqrt{\frac{\alpha}{8}} \ M_{Planck}$. 

The fields $S$ and $T_i$ are very familiar from previous studies,
especially for heterotic strings, whereas the fields $M_\alpha$ are
less well known. Even though these fields are attached to particular fixed
points (twisted moduli) they may play an interesting role on breaking
 supersymmetry, therefore we will include them in our discussion of
soft breaking terms. In order to do that, we need to consider the
expression for the Kahler potential as a function of $S,T_i$ and
$M_\alpha$. To simplify matters we will concentrate on a single overall
modulus $T$ and one blowing-up mode $M$.

General arguments and explicit calculations have been used to write
the K\"ahler potential for these fields
\cite{afiv}. At tree level it takes the
form
\be \label{kpot}
K =-\log(S+\bar S)-3\log(T+\bar T)
      +\sum_\alpha{C_\alpha\bar C_\alpha\over T+\bar T}
      +\hat{K}(M,\bar M; T, \bar T)\, ,
\ee
where $C_\alpha$ represent the matter fields of the theory. 
We have left the dependence on the field $M$ general, with 
the understanding that $\hat{K}$ mixes the fields $T$ and $M$. 

In addition there are usually
anomalous $U(1)$ symmetries which induce Fayet-Iliopoulos terms
of the form $\xi\sim \hat{K}_M$. These (in contrast to
the heterotic string case) 
may consistently be zero, giving mass to the $U(1)$ gauge
field but allowing the possibility of that symmetry remaining as a
global symmetry \cite{poppitz,iru,iq}.

\section{Soft Supersymmetry  Breaking Terms}

In this section we will present a  general analysis of soft
supersymmetry breaking terms in a class of models which admit an
intermediate fundamental scale. As we mentioned above, we
 will concentrate on orientifold\footnote{Analogous 
results are expected in the compact type IIB orbifold (not-orientifold) 
examples discussed in ref.\protect\cite{aiqu}.}
compactifications of type IIB strings. These models have been 
subject to intense investigation recently and a number of interesting
results have been obtained that allow us to study the structure of soft
breaking terms. They share some similarities 
with the better known soft terms derived from
heterotic string compactifications \cite{bim} under the
simple assumption of dilaton/moduli dominance of the
origin of SUSY-breaking, but there are also 
some important differences.

These compactifications have generically three different classes of
moduli-like fields. 
As in the perturbative heterotic string, we have 
the standard dilaton field $S$ and the moduli fields
of which we will consider, for simplicity, only an overall modulus $T$
that measures the overall size of the compact space. 
As discussed in the previous section,
there are also Ramond-Ramond fields which are not present in 
perturbative heterotic models. Again for simplicity
we will consider only a single one of these which we call $M$.
The $S$ and $T$ fields propagate in the whole of space-time, \ie the bulk,
whereas the $M$ fields are localized in a particular brane\footnote{Recently
the $M$ fields were considered as a source of 
gaugino masses after supersymmetry breaking in \cite{benakli2}.}.
 However, the 
$M$ fields can still 
play an important role in supersymmetry breaking,
in the sense that its $F$-term may get  a non vanishing vacuum expectation
value (VEV)
breaking supersymmetry, if the $M$ field lives in the observable brane
this will induce supersymmetry breaking on the brane, otherwise the
breaking may be communicated by the couplings to the $S$ and $T$
fields.
We will then 
analyze the structure of soft breaking terms when all three fields
$S$, $T$ and $M$ can have non-vanishing $F$-terms and therefore contribute
to supersymmetry breaking.
We will also consider matter fields $C_\alpha$, and will assume that the
full gauge group of the standard model comes from one single brane, which
is the most generic case in the explicit models studied so far
and is the natural scenario for gauge coupling unification.

We now proceed to compute the soft breaking terms  using the
tree-level K\"ahler potential (\ref{kpot}). 
As usual, for $F$-term
breaking we need to consider the $F$ part of the scalar potential
\beq
V\ =\ e^G\left(G_a(G^{-1})^a_bG^b-3\right),
\eeq
where $G=K+\log|W|^2$ and $W$ is the, unspecified, superpotential.
We will closely follow the analysis of ref.\cite{bim}
defining different goldstino angles for the mixing between the $F$
terms of the fields $S$, $T$ and $M$
and assume that SUSY-breaking effects are dominated by them.
 The feature here 
that differs from the previous approaches 
is the inclusion of the field $M$
as another source for supersymmetry breaking. Also, in contrast to the
heterotic string, there is no self $T$-duality in these models and
therefore there are no $T$-dependent holomorphic threshold corrections to
the gauge couplings.

In this class of models there is a mixing between the $T$ and $M$ fields
due to  the presence of  a generalized Green-Schwarz mechanism
and the function $\hat{K}$ of (\ref{kpot}) is of the form
\be \label{khat}
\hat{K}=\hat{K}(M+\bar M-\dgs\log(T+\bar T)).
\ee
Thus one has the K\"ahler metric for $S$, $T$, $M$ ($m,n=S,T,M$, $i,j=T,M$):
\be \label{kmet}
(K_{m\bar n})=\left(\begin{array}{cc}\displaystyle{1\over(S+\bar S)^2} &0\\
                    0
&\displaystyle(K_{i\bar j})\end{array}\right)\,,\quad\w
(K_{i\bar j})=\left(\begin{array}{cc}\displaystyle
       K_{T\bar T}  &\displaystyle{-\dgs\hat{K}''\over T+\bar T}\\
       \displaystyle{-\dgs\hat{K}''\over T+\bar T} &\displaystyle \hat{K}''
                   \end{array}\right)\,,
\ee
where
$$K_{T\bar T}={1\over(T+\bar T)^2}
              \left(3+2\sum_\alpha {C_\alpha\bar C_\alpha\over T+\bar T}
                    +\dgs^{\;2}\hat{K}''+\dgs\hat{K}'\right)$$
and the primes denote derivatives with respect to the argument of $\hat{K}$.

Since there is mixing between the $T$ and $M$ fields,
we would like to normalize the kinetic terms, \ie multiply by a matrix $P$
such that $P^\dagger(K_{i\bar j})P=I$.
Defining\footnote{Note that in limit where all fields are at the minimum of
their potential, \ie $C_{\alpha}=0=\hat{K}'$, one has $k=3/\hat{K}''$.}
$k=(T+\bar T)^2 K_{T\bar T}/\hat{K}''-\dgs^{\;2}$
and expanding in powers of $1/(T+\bar T)$ we obtain
\be
    P={1\over\sqrt{\hat{K}''}}\left(\begin{array}{cc}
      {{T+\bar T}\over\sqrt{k}}  &-{\dgs\over T+\bar T}\\
      {\dgs\over\sqrt{k}} + {\dgs\sqrt{k}\over(T+\bar T)^2}  
               &\ 1-{\dgs^{\;2}\over(T+\bar T)^2}
     \end{array}\right)
    +\cO\left({1\over(T+\bar T)^3}\right)\,.
\ee
In this expansion we assumed that $\hat{K}''$ goes to a constant in
the limit $T+\bar T\to\infty$.

After having diagonalised the K\"ahler metric we can now
define the mixing between the $F$-terms of the fields $T$ and $M$
by using a vector $\Theta$ of modulus one:
\be
\Theta=
\left(\begin{array}{c}\Theta_{\bar T}\\ \Theta_{\bar M}\end{array}\right)=
\left(\begin{array}{c}\sin\phi\\ \cos\phi\end{array}\right)\,.
\ee
The $F$-terms $F^T$ and $F^M$ are then defined by
\be
\left(\begin{array}{c}F^T\\F^M\end{array}\right)\  =\  \sqrt3 C\mgr\cos\theta
  P\Theta 
\ee
where $C=\sqrt{1+{V_0\over3\mgr^2}}$
and $V_0$ is the vacuum energy.

We can now write an explicit expression for  the corresponding F-terms
in the large $T$ limit:
   
\bea
F^T &\approx &\sqrt3C\mgr\cos\theta\left({{T+\bar T}\over\sqrt{k}}\sin\phi
        -{\dgs\over T+\bar T}\cos\phi  \right)\,, \nonumber\\
F^M & \approx &\sqrt3C\mgr\cos\theta\left(
 \left({\dgs\over\sqrt{k}} + {\dgs\sqrt{k}\over(T+\bar T)^2}\right)\sin\phi
          +\left(1-{\dgs^{\;2}\over(T+\bar T)^2}\right)\cos\phi\right)\,.
\eea

For the dilaton $S$ there is no mixing and its F-term is simply given by
\be
F^S=\sqrt3C\mgr\sin\theta(S+\bar S)\,.
\ee

\subsection{Gaugino masses}
\medskip

In general the gaugino masses for gauge group $G_a$ are given by
\be \label{gmass}
M_a=\half(\R f_a)^{-1}F^m\partial_m f_a\,.
\ee
Here $f_a=S+{s_a\over4\pi}M\,$ denotes the gauge kinetic function
normalised as $\R f_a=4\pi/g_a^2$. We then find
\be
M_a  = {\sqrt3C\mgr\over S+\bar S+{s_a\over4\pi}(M+\bar M)}
 \left(\sin\theta(S+\bar S)+{s_a\over4\pi}
  \cos\theta(P_{M\bar T}\sin\phi+P_{M\bar M}\cos\phi)\right)\,.
\ee
In the limit $T+\bar T\to\infty$ and using
$2\R f_a=2\,\alpha_a^{-1}$ and $S+\bar S=2\,\agut^{-1}$
we then obtain
\be 
 M_a=\sqrt3C\mgr{\alpha_a\over\agut}\left(\sin\theta
      -{s_a\over8\pi}\agut\,\cos\theta\left(
            {\dgs\over\sqrt{k}}\sin\phi + \cos\phi\right)
      \right)+\cO\left({1\over(T+\bar T)^2}\right)\,.
\ee

\medskip

\subsection{Scalar masses}
\medskip
Writing the K\"ahler potential (\ref{kpot}) as
\be  \label{kdecom}
     K=\bar K(S,\bar S,T,\bar T,M,\bar M)
      +\sum_\alpha \tilde K_\alpha(S,\bar S,T,\bar T,M,\bar M)
            C_\alpha\bar C_\alpha\,,
\ee
the scalar mass squared of the field $C_\alpha$ is given by
\be m_\alpha^2=(\mgr^2+V_0)-\overline{F^m}
				F^n\partial_{\bar m}\partial_n
                             \log\tilde K_\alpha\,.
\ee
In our case we find
\be
m_\alpha^2=\mgr^2\left(1-3C^2\cos^2\theta{(P_{T\bar T}\sin\phi
              +P_{T\bar M}\cos\phi)^2\over(T+\bar T)^2}\right)+V_0\,.
\ee
In the limit $T+\bar T\to\infty$ we obtain  
\be
m_\alpha^2=V_0+\mgr^2\left(1-\frac{3}{k}
           C^2\cos^2\theta\sin^2\phi\right)
           +\cO\left({1\over(T+\bar T)^2}\right)\,.
\ee

\medskip

\subsection{A-terms}
\medskip

The A-terms are derived from the formula
\be A_{\alpha\beta\gamma}=F^m\left(\bar K_m+\partial_m\log
        Y_{\alpha\beta\gamma}-\partial_m\log(\tilde K_\alpha\tilde K_\beta
        \tilde K_\gamma)\right)\,.  
\ee
{}From the structure of the K\"ahler potential
(\ref{kdecom}) and assuming
 that the Yukawa couplings do not depend on the moduli, \ie
$\partial_m\log Y_{\alpha\beta\gamma}=0$, we obtain
the following in the large-volume limit:
\beqa
A_{\alpha\beta\gamma} &=& -\sqrt3C
\mgr\bigg(\sin\theta+\cos\theta\cos\phi\hat{K}'
         \bigg)+\cO\left({1\over(T+\bar T)^2}\right)\,.
\eeqa

\subsection{Dilaton Domination Scenario}

Rather than discussing details of the general scenario
for the soft breaking terms, it is more
useful to concentrate on particular limiting cases which tend to have very
different physical implications.
Let us begin therefore by considering
the well studied dilaton domination scenario which can be
obtained by setting $\cos \theta=0$. We find
\beqa
M_a & = & \sqrt{3}\ m_{3/2} (1+\kappa_a)^{-1}, \qquad
\kappa_a=\frac{s_a}{4\pi} \frac{M+\bar M}{S+\bar S}\nonumber\\
m_\alpha^2 & = & m_{3/2}^2\nonumber\\
A & = & -\sqrt{3}\ m_{3/2}\, ,
\label{dildom}
\eeqa
where we have set $C=1$ (vanishing cosmological constant).
Notice that the $\kappa_a$ dependence makes the gaugino masses
non universal, in contrast to the heterotic case.  
This, however, is a one-loop effect and to first approximation,
for very small values of $\kappa_a$,
the dilaton dominated type I string has the same
supersymmetry breaking terms as the standard dilaton
dominated perturbative heterotic string.
As we shall see in the next section, the big
difference arises when we start the RG running at the
intermediate scale, something which is allowed for type I strings but
it not for perturbative heterotic string.

\subsection{$M$ Domination Scenario}
 A completely new scenario which is allowed in these models
is the limit in which only the
$M$ field is the main source of supersymmetry breaking.
We obtain this scenario by setting $\sin\theta=\sin\phi=0$ in the
 above equations, and we find the following expressions for the soft
 breaking terms:
\beqa
M_a & = & -\frac{\sqrt{3}}{8\pi}\alpha_as_a \ m_{3/2}\nonumber\\
m_\alpha^2 & = & m_{3/2}^2 \nonumber\\
A & = & -\sqrt{3}\ m_{3/2} \hat{K}' \,.
\eeqa

This scenario applies only if the $M$ field lives in the same
brane as the standard model fields. $M$ domination
could also occur with the $M$ fields and standard model
fields living on different branes,
however in that case the induced soft breaking terms will all
be vanishingly small since $M$ does not couple directly to the
visible sector.

\subsection{Moduli Domination Scenario}
The $T$ dominated scenario corresponds to the limit
$\sin\theta=\cos\phi=0$, for which we obtain
\beqa
M_a & = & -\frac{\sqrt{3}\,\dgs}{\sqrt{k}}
\frac{s_a\alpha_a}{8\pi}\ m_{3/2}\nonumber\\
 m_\alpha^2 & = & m_{3/2}^2 \frac{k-3}{k}
\nonumber\\
A & = &0.
\eeqa

Notice that the structure of the soft breaking terms is different from
that found in ref.\cite{bim}\ for the moduli domination scenario in
heterotic models. The main reason for this difference is the fact that
in the heterotic case there are $T$ dependent threshold corrections
for the gauge couplings which are required by $T$-duality. Type I models
are not self dual under $T$ duality and there are no $T$ dependent thresholds
\cite{iru}.

Notice that the moduli and $M$ domination soft breaking terms
simplify substantially if the minimum of the $M$ field potential
is at vanishing Fayet-Iliopoulos term for which $\hat{K}'=0$. In this
case one has $k=3$ if the VEVs of the charged fields vanish.
Independent of this, in both scenarios the gaugino masses are one-loop
suppressed, unlike the dilaton dominated case. Therefore
 these soft terms are of the
same order as or dominated by those induced by anomaly mediated
supersymmetry breaking~\cite{amsb}.
Hence anomaly mediation should also be included in the analysis
and cannot be neglected in the renormalization group analysis.
Furthermore, once one-loop effects become relevant, we would have to
consider
loop corrections to the K\"ahler potential which are 
 not yet well
understood in type I models.
 A complete analysis including the effects
of anomaly mediated supersymmetry breaking is beyond the scope of the
present article and it is left for a future publication. In the
following we will therefore concentrate on the dilaton dominated scenario
in which the effects of anomaly mediation are negligible.

\section{Phenomenological Considerations}

The phenomenological study of SUSY breaking in various string models has
in the past been almost exclusively 
focused upon high (GUT or higher) scale stringy 
models~\cite{stringpheno,sakisf}.
Here, we turn to pertinent phenomenological consequences of the dilaton
domination scenario
discussed above. We perform the analysis for an intermediate string scale
$M_X=10^{11}$ GeV {\em and}\/ the 
conventional GUT scale $M_{GUT}=2 \times 10^{16}$ GeV. This allows us to
contrast the intermediate scale models with their GUT-scale counterparts.
We will focus upon the case where the effective theory
below the string scale is the MSSM\@. 
The effective theory below the string scale affects the results by altering
the renormalization of the 
soft SUSY breaking and supersymmetric parameters down to the weak scale.
In fact, as we demonstrate, going beyond
the MSSM approximation by adding extra leptons in order to achieve gauge
unification does not significantly alter many of the results. Where there is a
large change, we display the results when we take new leptons into account
in the gauge beta functions.

First, we show that adding just a few (in an explicit example, 5) extra
vector-like representations of leptons can achieve gauge unification at the
intermediate scale.
We then perform an analysis of Yukawa unification. The implications and
viability of bottom-tau Yukawa unification 
are investigated
briefly. These results are approximately independent of the assumed form of
SUSY breaking, so we ignore sparticle splittings in order to make the analysis
more model independent. For the other phenomenological analyses, we take the
full non-degenerate sparticle spectrum into account.
Next, the assumed limit of SUSY breaking (dilaton dominated) is 
employed in order to set the boundary conditions of
the soft SUSY breaking parameters at the string scale.
We will focus upon the spectra, charge and colour breaking (CCB) bounds and
fine-tuning measure in each case.
We then combine these quantities with the
experimental bounds upon sparticle masses to identify the allowed parameter
space, how fine-tuned it
is, and predict the sparticle and Higgs masses. This will allow a useful
comparison of the two different string-scale scenarios.

\subsection{Gauge and Yukawa Unification}

We now turn to the constraints and fits from gauge and Yukawa unification,
both of which may be successful in SUSY GUTs~\cite{susyguts}. In this
subsection only, we will use the two-loop MSSM RGEs~\cite{MandV} above
$m_t$, thus assuming a degenerate MSSM spectrum (with the extra states) at
$m_t$. Later, when we  consider
SUSY breaking phenomenology, we will therefore go beyond this approximation,
taking sparticle splittings into account. 
Although loop corrections  involving sparticles to the 
weak-scale Yukawa couplings are expected, we will ignore them. This 
approximation allows us to make broad statements that do not depend upon the
details of SUSY breaking and are therefore more model independent. 

The issue of gauge unification at the intermediate scale has already been
studied to one-loop order~\cite{biq}. It was suggested that extra
leptons may be added to the MSSM in order to achieve it. The simplest model
found involved adding $4 (L_L + E_R)$ supermultiplets in vector-like copies to
the MSSM\@. 
We now examine this
statement to two loop order, in the hope of finding the minimal addition of
leptons to the MSSM which achieves gauge unification at the intermediate
scale. To two-loop order and using central experimental inputs 
for the gauge couplings and masses as in section~\ref{sec:NA}, we find that 
$2 \times L_L + 3 \times E_R$ extra vector-like
representations\footnote{Note that we assume the extra states do not
contribute to electroweak symmetry breaking, or mix with the MSSM leptons.}
are enough to achieve approximate gauge unification at $M_X
\sim 10^{11}$ GeV. With this spectrum, we obtain
\beq
g_1(M_X) = 0.81, \qquad g_2(M_X) = 0.82, \qquad g_3(M_X) = 0.81.
\eeq
As a case study, we will consider the MSSM augmented by $2 \times L_L + 3
\times 
E_R$ vector-like
representations
when we want to examine the 
possible effect 
of adding extra states to the MSSM spectrum in order to
achieve gauge unification. The default analysis will however be valid for the
MSSM, where we assume that either the corrections from the extra states are
small, 
or that mirage unification (where stringy corrections change the boundary
conditions at the string scale in just the correct way to agree with the
measured gauge couplings)
occurs at $M_X$.

Yukawa unification in the third family can be a prediction of SUSY GUTs, and
has successfully passed empirical constraints~\cite{yukun, largecor}. It may
also be 
predicted  in some particular 
string models. This prediction may appear  in either its weaker form 
\beq R_{b/\tau}\equiv \frac{h_b(M_X)}{h_\tau(M_X)}=1 \label{btau} \eeq
or in a stronger form 
\beq h_t(M_X)=h_b(M_X)=h_\tau(M_X). \label{tripleyuk} \eeq
We do not analyse the stronger form because it requires the high $\tan \beta$
region requires the addition of large finite corrections, which are not
available to us at this time. In the canonical GUT scale universal scenario,
these finite corrections render Eq.~\ref{tripleyuk} incompatible with the
data~\cite{largecor}.

Instead, we turn to the more general constraint in Eq.~\ref{btau}.
For a given $\tan \beta$, we determine the third family Yukawa couplings
at $m_t$ as in section~\ref{sec:NA} by running the
low energy empirical inputs up. 
\FIGURE[t]{%
\epsfig{file=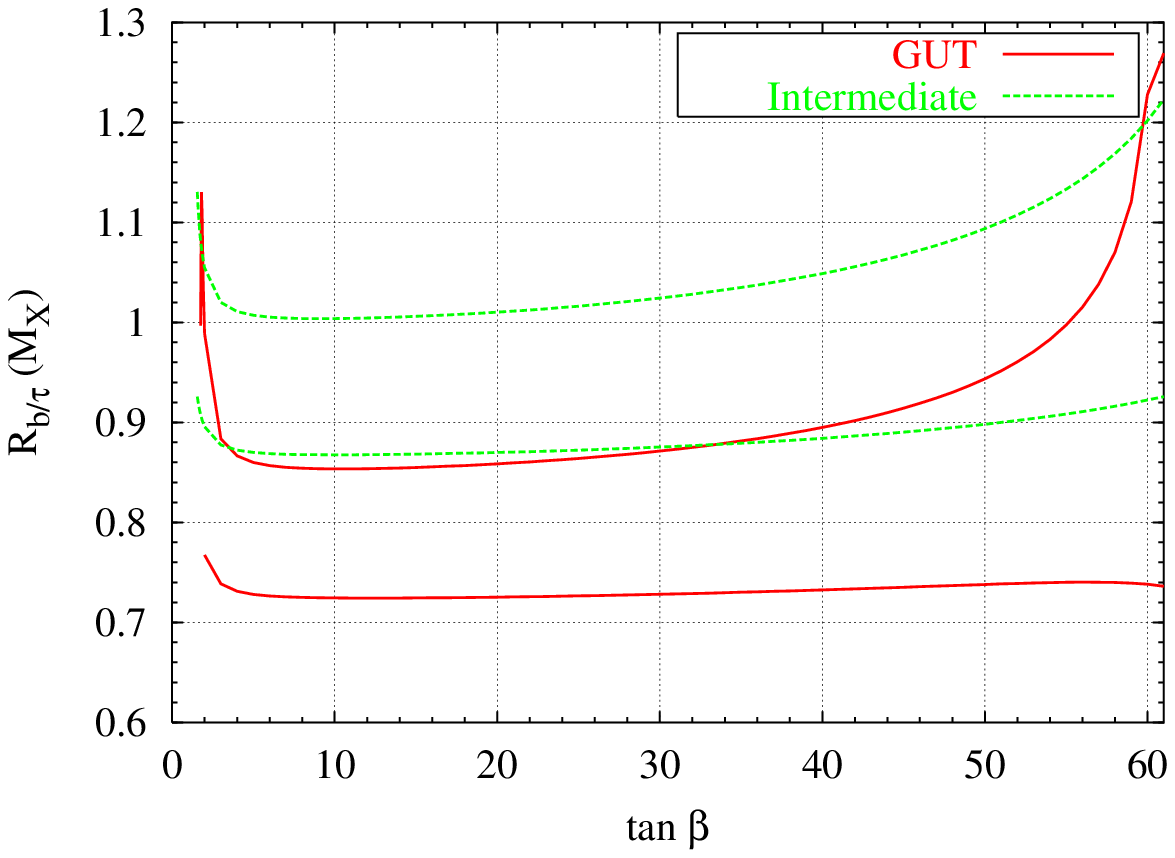, width=5in}
\label{fig:btau}
\caption{Bottom-tau Yukawa unification in the GUT and intermediate
scenarios, where $M_X=2 \times 10^{16}$ GeV and $10^{11}$ GeV respectively.}
}
We run the Yukawa and
gauge couplings (using the MSSM RGEs) from
$m_t$ to $M_X$ for $M_X=2 \times 10^{16}$,$10^{11}$ GeV, \ie the GUT-scale
and intermediate scale unification hypotheses respectively.

$R_{b/\tau}$, as defined in eq.(\ref{btau}), is displayed for the intermediate
and GUT-scale unification scenarios for various $\tan \beta$ in
fig.~\ref{fig:btau}. For each scenario, the range of $R_{b/\tau}$ predicted by
varying $\alpha_s=0.119 \pm 0.002$, $m_\tau(m_\tau)=1.77705\pm 0.00027$ GeV,
$m_b(m_b)=4.25 \pm 0.15$ GeV and
$m_t(m_t)=165 \pm 5$ GeV within their 1$\sigma$ errors is depicted as the
region between two lines.
In the MSSM SUSY GUT scenario (\eg SUSY minimal $SU(5)$), this region (between
the two solid lines)
constrains $\tan \beta$ to be either $2-3$, 
or\footnote{We note that there are large uncertainties in the large $\tan
\beta>40$ region from neglected finite corrections\cite{largecor}, and as
such our results may not be quantitatively accurate there.}
$\geqsim 55$, since that is
where it crosses $R_{b/\tau}=1$.
In the intermediate scale scenario, we see that there is no such constraint
upon $\tan \beta$ because there is a point between the dashed lines consistent
with $R_{b/\tau}=1$ for any $\tan \beta$.

\subsection{CCB bounds}

Supersymmetric models have many directions in field space 
that are $F$ and/or $D$ flat,
and supersymmetry breaking can cause these directions 
to develop global minima which break charge and colour (CCB
minima)~\cite{ccb1,casas1,dilaton,ccb2,as98,casas2,ccb3,sandb}, and we
shall include the constraints that arise from avoiding such minima
in our later analysis. Of particular relevance 
is the fact that there is always such a minimum in 
dilaton dominated models if unification occurs at the GUT 
scale~\cite{dilaton}. 
(These minima also afflict M theory models~\cite{as98,casas2}.) 
This is a pity since dilaton domination has many phenomenological 
advantages in addition to the purely aesthetic ones of simplicity 
and predictability. For example, dilaton dominance guarantees universal
supersymmetry breaking terms thereby solving the problem of 
large FCNCs. 

Before continuing we should mention a 
general `solution' to the problem of CCBs which is really a 
cosmological observation; the rate of tunneling from a false 
vacuum (\ie the physical vacuum in which we are living) to the 
global CCB minimum is usually many orders of magnitude longer than the
age of the universe in standard cosmology. Thus provided that cosmology
can also place the universe in the relatively small 
physical vacuum initially
(for example during a period of heating to temperature higher than the
supersymmetry breaking scale) the existence of a CCB minimum
is allowed. On evidence,
since dilaton domination has fallen out of favour,
this solution does not seem to be very appealing. 

In our numerical results we shall see
that one of the most significant effects of  reducing the 
string scale is a complete change in the behaviour of 
CCB minima. In fact at lower string scales 
the most restrictive CCB minima disappear and dilaton domination 
is allowed once again. We find this solution to CCB minima
an appealing feature of a lower string scale. 
This change in the CCB bounds was anticipated near the 
low $\tan\beta $ quasi-fixed point (where the Yukawa 
coupling blows up at the {\em string}\/ scale) in 
ref.\cite{sandb} by using approximate analytic 
solutions for the renormalization group equations.
Since that type of analysis is necessarily rather technical 
we briefly summarize the main findings.

There are two important kinds of bounds; those corresponding to 
$D$-flat directions which develop a minimum due to large trilinear 
supersymmetry breaking terms; those corresponding 
to $D$ {\em and} $F$ flat directions which 
correspond to a combination of gauge invariants involving $H_2$. 
The first kind of flat directions give a familiar 
set of constraints on the trilinear couplings which is typically of the form 
\be 
A_t^2 \leqsim 3 (m_{H_2}^2 + m_{t_R}^2 + m_{t_L}^2 ),
\ee
where the notation is conventional. These constraints turn out to 
be very weak. Much more severe bounds come from the directions which are 
$F$ and $D$ flat. (The bound can
be optimized as in ref.\cite{casas1} and indeed as has been done in
our numerical analysis, but the optimal direction is very close to the
$F$ and $D$ flat direction and the bounds do not change significantly.) 
 
$F$ and $D$ flat directions can be
constructed from conjunctions of $LH_2$ plus any one of the following 
gauge invariants~\cite{as98}, 
\be 
LLE\mbox{, }
LQD\mbox{, }
QULE\mbox{, }
QUQD\mbox{, }
QQQLLLE.
\ee
Absence of CCB minima along the first two directions is usually 
enough to guarantee their absence along the rest~\cite{as98}.
As an example consider the $L_iL_3E_3$, $L_iH_2$ direction, which 
corresponds to the choice of VEVs,
\bea
\label{komkom}
h_2^0             &=& -a^2 \mu/h_{E_{33}} \nonumber \\
\tilde{e}_{L_3}=\tilde{e}_{R_3} &=& a \mu/h_{E_{33}} \nonumber \\
\tilde{\nu}_i   &=& a  \sqrt{1+a^2} \mu/h_{E_{33}},
\eea 
where $a$ parameterizes the distance along the flat direction.
The potential along this direction
depends only on the soft supersymmetry breaking terms;
\be
\label{softv}
V=\frac{\mu ^2}{h_{D_{33}}^2} a^2 (a^2 (m_{H_2}^2+m_{L_{ii}}^2) + 
m_{L_{ii}}^2+m_{E_{33}}^2+m_{L_{33}}^2 ).
\ee 
In order to minimize the one-loop corrections, the mass squared parameters 
in eq.(\ref{softv}) are evaluated at a renormalization
scale of $Q=\mbox{max}(h_t h_2^0,M_{susy})$. 
The first term in the potential dominates at large VEVs when $a\gg 1$ and, 
because $m_{H_2}^2<0$ in order to give electroweak symmetry
breaking, this radiatively generates a dangerous CCB minimum with
a VEV which is typically a few orders of magnitude larger 
than the weak scale.

The general weakening of the CCB bounds as we lower the string scale 
is caused by the interplay of the first ($a^4$) and second ($a^2$)
terms in eq.(\ref{softv}). Both terms are assumed to be positive
at the string scale but only the first can become negative 
due to the strong scale dependence of $m_{H_2}^2$. However when
$a\ll 1$ the second term dominates 
since the mass squared terms are of the same
order. Hence if a CCB minimum forms at
all along this direction it can only do so for $a\gg1$ or from 
eq.(\ref{komkom})
\be
h_2^0 \gg \left|\frac{\mu}{h_{E_{33}}}\right|.
\ee
This implies that if we choose a string scale 
which is less than $ \left|\frac{
\mu}{h_{E_{33}}}\right|$ the possibility of a CCB minimum along this
direction is excluded entirely. The behaviour of the 
bounds as the unification scale is increased towards the usual GUT scale
was examined in ref.\cite{sandb}. This entailed a 
detailed treatment of the renormalization group equations, but the 
net result is that the CCB bounds increase rather smoothly 
towards their usual GUT scale values. 

The $F$ and $D$ flat direction discussed here 
provides the severest bounds on the 
parameter space and for the usual constrained MSSM with 
$M_X=2\times 10^{16}$ it can be expressed 
as a bound on the degenerate scalar mass at the GUT scale,
\be 
m_0 < \lambda(\tan\beta) M_{a} ,
\ee
where the GUT scale gaugino masses are of course degenerate, and 
where $\lambda(\tan\beta) \approx 1$ at the quasi-fixed point
and falls off to $\approx 0.4$ and larger values of $\tan\beta$. 
The $m_0$ and $M_{a}$ parameters are related by the dilaton and 
moduli VEVs, and we can immediately see that the dilaton dominated
scenario is ruled out by the above. We show this numerically later
using the techniques summarized in ref.\cite{casas1}

\subsection{Fine Tuning}

At tree-level, the $Z$ boson mass is determined to be 
\begin{equation}
\frac{1}{2} M_Z^2 = \frac{m_{H_1}^2 - m_{H_2}^2 \tan^2 \beta}{\tan^2 \beta -
1} - \mu^2 \label{FTtree}
\end{equation}
by minimizing the Higgs potential. $\tan \beta$ refers to the ratio of Higgs
vacuum 
expectation values (VEVs) $v_1/v_2$ and $\mu$ to the Higgs mass parameter in
the MSSM superpotential. 
In the universal models discussed here, $m_{H_2}$ has the same origin as the
super-partner masses ($m_0$). Thus as
search limits put lower bounds upon super-partners' masses, the lower bound
upon $m_0$ rises, and consequently so does $|m_{H_2}|$. A cancellation is then
required between the first and second terms of eq.(\ref{FTtree}) in order
to provide the measured value of $M_Z \ll |m_{H_2}|$. Various measures have
been proposed in order to quantify this cancellation~\cite{measures}.

The definition of naturalness $c_a$ of a `fundamental' parameter $a$ employed
in ref.\cite{focuspoints} is
\begin{equation}
c_a \equiv \left| \frac{\partial \ln M_Z^2}{\partial \ln a} \right|.
\end{equation}
From a choice of a set of fundamental parameters $\{ a_i \}$, the
fine-tuning of a particular model is defined to be $c \equiv \mbox{max}(c_a)$.
Our choice of free, continuously valued, independent and fundamental
parameters are 
\begin{equation}
a_i \in \{ \mu,\ B,\ m_{3/2} \},
\end{equation}
\ie the one on the right hand side of eq.(\ref{dildom}), but
augmented by the superpotential Higgs mass $\mu$ and the soft SUSY breaking
Higgs bilinear
parameter $B$. Notice that, following ref.\cite{focuspoints}, we have not
explicitly considered variations of $M_Z$ with respect to the top Yukawa
coupling $h_t$.
The origin of soft terms and Yukawa couplings are so different
that putting them on equal footing seems, in our opinion,   inadequate.
 Other authors have considered including
$h_t(M_X)$~\cite{rs}, in scenarios where it makes a large difference ($m_0 \gg
M_{1/2}$). 
We have checked that here, the exclusion or inclusion of $h_t$
makes no difference since the fine-tuning measure comes 
predominantly from $\mu$.

\subsection{Gaugino masses}

It is possible to make some predictions of the MSSM gaugino masses
$M_{1,2,3}$ at low energy scales depending upon their boundary conditions at
the unification scale $M_X$.
We are able to do this because to one-loop order and for 
the MSSM (with any additional extra $N=1$ supermultiplets), we
have~\cite{MandV}
\beq
16 \pi^2 \frac{d (M_i / \alpha_i)}{d \ln \mu} = 0 \Rightarrow
\frac{M_i (\mu)}{\alpha_i (\mu)} = \frac{M_i(M_X)}{\alpha_i(M_X)}.
\eeq
In the standard unification scenario where gauge couplings {\em and}\/ gaugino
masses are unified at $M_X$, this leads to the familiar
prediction 
\beq
M_1(\mu) \alpha_1(\mu)^{-1} = M_2(\mu) \alpha_2(\mu)^{-1} = M_3(\mu)
\alpha_3(\mu)^{-1},
\eeq
valid at any renormalisation scale $\mu$. In particular, we may take $\mu
\approx M_Z$ in order to estimate $M_{1,2,3}$. From experiment, we have
$\alpha_{1,2,3}^{-1}(M_Z) \sim \{ 90,30,8\}$ roughly (for $\alpha_1$ in the
GUT normalisation). Thus
\beq
9 M_1 \sim 3 M_2 \sim M_3 \label{usual}
\eeq
predicts the gaugino mass ratios in the canonical unification scenario, $M_X
\sim 10^{16}$ GeV.
If $M_{1,2,3}$ are deduced from  experiments, the relation eq.(\ref{usual})
would therefore provide an immediate test of this scenario.
We note here that the same relation results in the case where the gauge
couplings meet at the intermediate scale (for example by adding the vector-like
copies of $2 \times E_R + 3 \times L$
to the MSSM).

\subsection{SUSY Breaking Numerical Analysis}
\label{sec:NA}
The softly broken MSSM RGEs used are contained within ref.\cite{MandV}.
We follow here a fairly standard numerical algorithm to calculate the MSSM
sparticle spectrum, \eg see ref.\cite{BBO}, so here we merely briefly review
our approximations.
The soft breaking parameters' RGEs are calculated to one-loop order with full
family dependence, but all
supersymmetric parameters are evolved to two-loop order. The Yukawa matrices
are approximated to be diagonal. 
To obtain their $\overline{MS}$ values (as well as that of $\alpha_s(\mu)$) at
$\mu=m_t(m_t)$, we use 3 loop
QCD$\otimes$1 loop QED as an effective theory to run the quark and lepton
masses up (with
step-function decoupling of quarks and the tau).

Near the weak scale, thresholds
from sparticles are modelled by the step function and all finite corrections
are neglected. Thus the sparticles are decoupled below their running mass as
in ref.\cite{sakisandlah}.
The tree-level MSSM Higgs potential supplemented by the largest (top
and stop) one-loop corrections~\cite{BBO} to the tadpoles is used to calculate
$\mu$ and $B$ from the radiative electroweak symmetry breaking constraints at
the scale where one-loop corrections are small~\cite{zwirneretal}, $\hat Q =
\sqrt{(m_{{\tilde t}_1}^2 + m_{{\tilde t}_2}^2)/2}$.
Tree-level mass matrices are used to calculate the sparticle masses from the
weak scale MSSM parameters, except for the Higgs masses, where 
state-of-the-art expressions including finite terms and some large two-loop
contributions are utilized~\cite{georgetc}.

\TABULAR[r]{cc}{\label{tab:inp}
$G_F=1.6639$ &  $M_Z=91.1867$ \\
 $M_W=80.405$ & $\alpha(M_Z)^{-1}=127.9$\\
$m_t(m_t)=160$ & $m_b(m_b)=4.25$\\
$m_\tau(m_\tau)=1.777$ & $\alpha_s(M_Z)=0.119$ \\
}{Inputs used in the
numerical analysis~\cite{PDG}.} 
The most important empirically-derived inputs used~\cite{PDG} are shown in
table~\ref{tab:inp}.
$m_x(\Lambda )$
are running masses in the $\overline{MS}$ scheme at scale $\Lambda $. 
The weak-boson masses are extracted in the on-shell scheme.
All masses are measured in GeV.
Note that $m_t(m_t)$ is kept constant over SUSY breaking parameter space. A
more accurate calculation would use a constant value of the empirically
derived pole mass, and derive the $\overline{MS}$ running mass from it by
taking QCD and gluino radiative corrections into effect.
Our procedure corresponds to taking a slightly different value of the top pole
mass at different points of parameter space, but should still be mostly within
$m_t=175 \pm 5$ GeV~\cite{PDG}. Similarly, $m_b(m_b)$ receives contributions
from sparticles for $\tan \beta \geqsim 40$ that we have neglected.

\TABULAR[r]{ccc}{\label{bounds}
$m_{\tilde g} > 300 $ & $m_{\tilde{t}_1} > 83 $ & $m_{\chi_1^0} >31.6 $ \\
$m_h > 89.3 $ & $m_{\chi_1^\pm} > 84$ & $m_{\tilde{e},\tilde{\mu}} > 80
$\\
$m_{\tilde{u},\tilde{d}} > 250$ &  $m_{\tilde{b}_1} >
83 $ &}
{Empirical lower bounds upon MSSM sparticle masses (in GeV)~\cite{sakisf}}
We will constrain the models to respect empirical lower bounds
upon MSSM particle masses~\cite{sakisf}, as shown in table~\ref{bounds}.
The most restrictive bounds turn out to be those upon $h^0$, the lightest CP
even Higgs, ${\tilde g}$ the gluino,  and $\chi_1^\pm$, the lightest charginos.

\subsection{Spectra}

We now turn to the spectrum of the 
dilaton dominated SUSY-breaking scenario, \ie using
universal boundary conditions as in eq.(\ref{dildom}). 
\FIGURE[t]{%
\fourgraphs{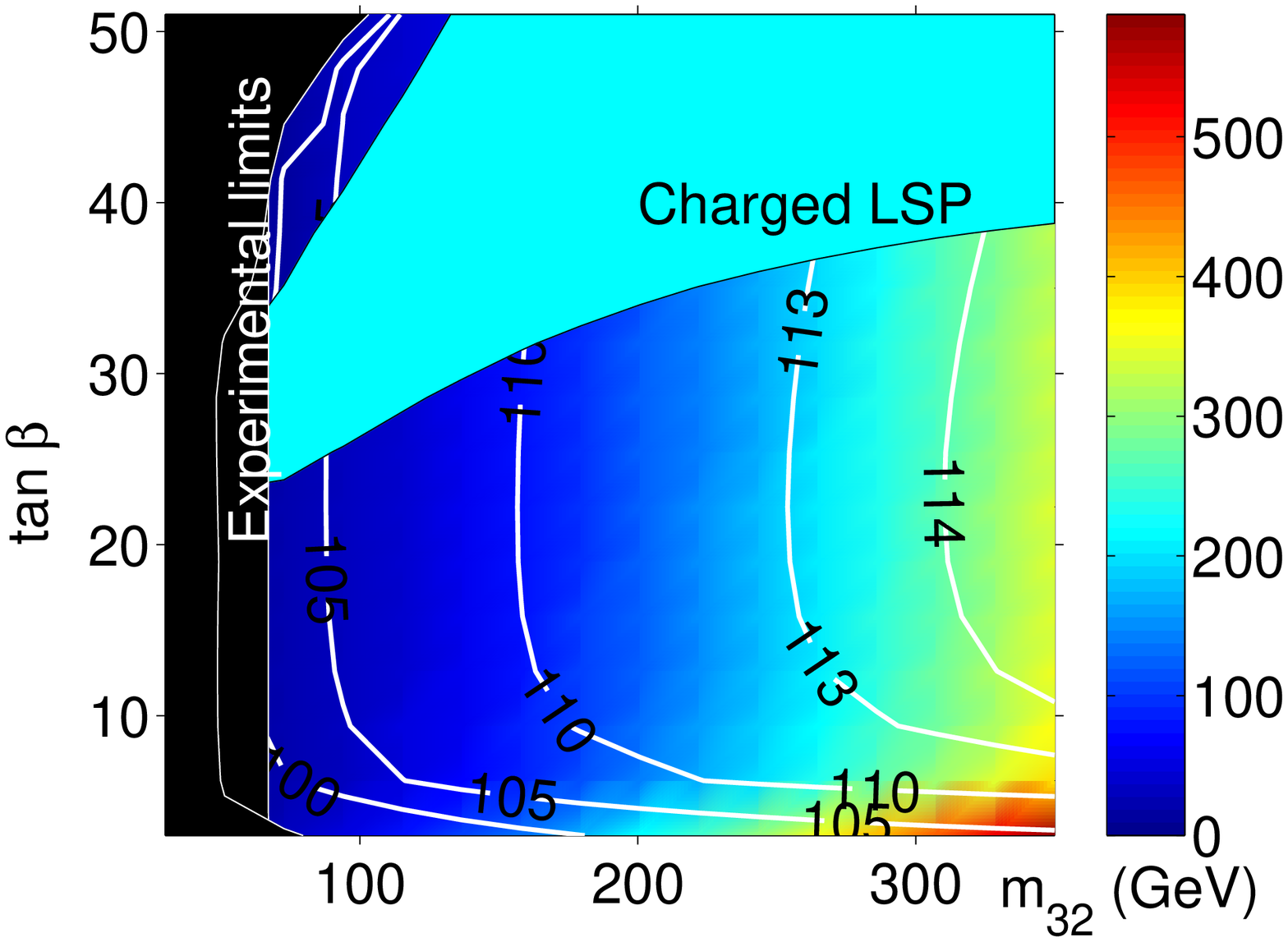}{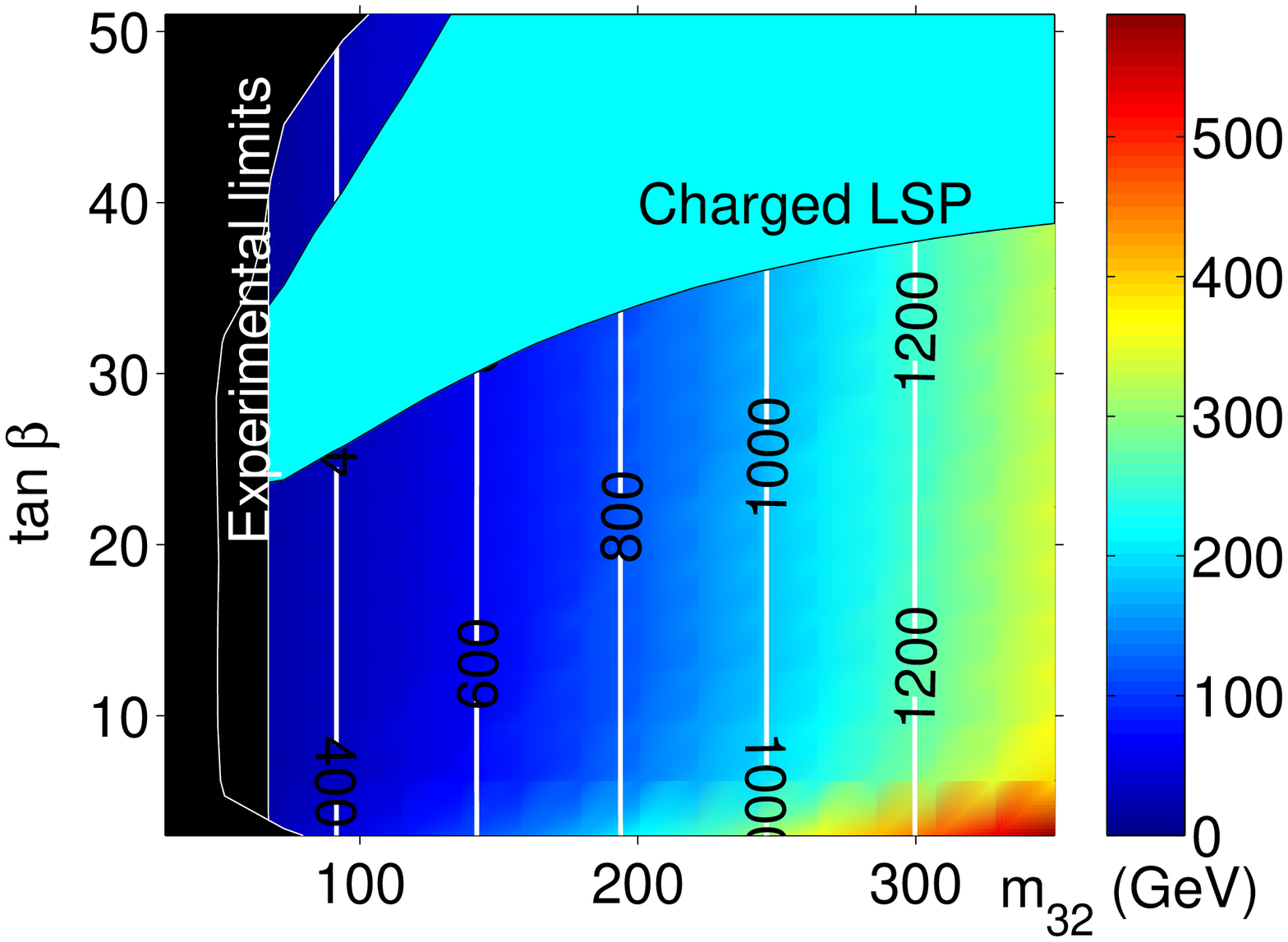}{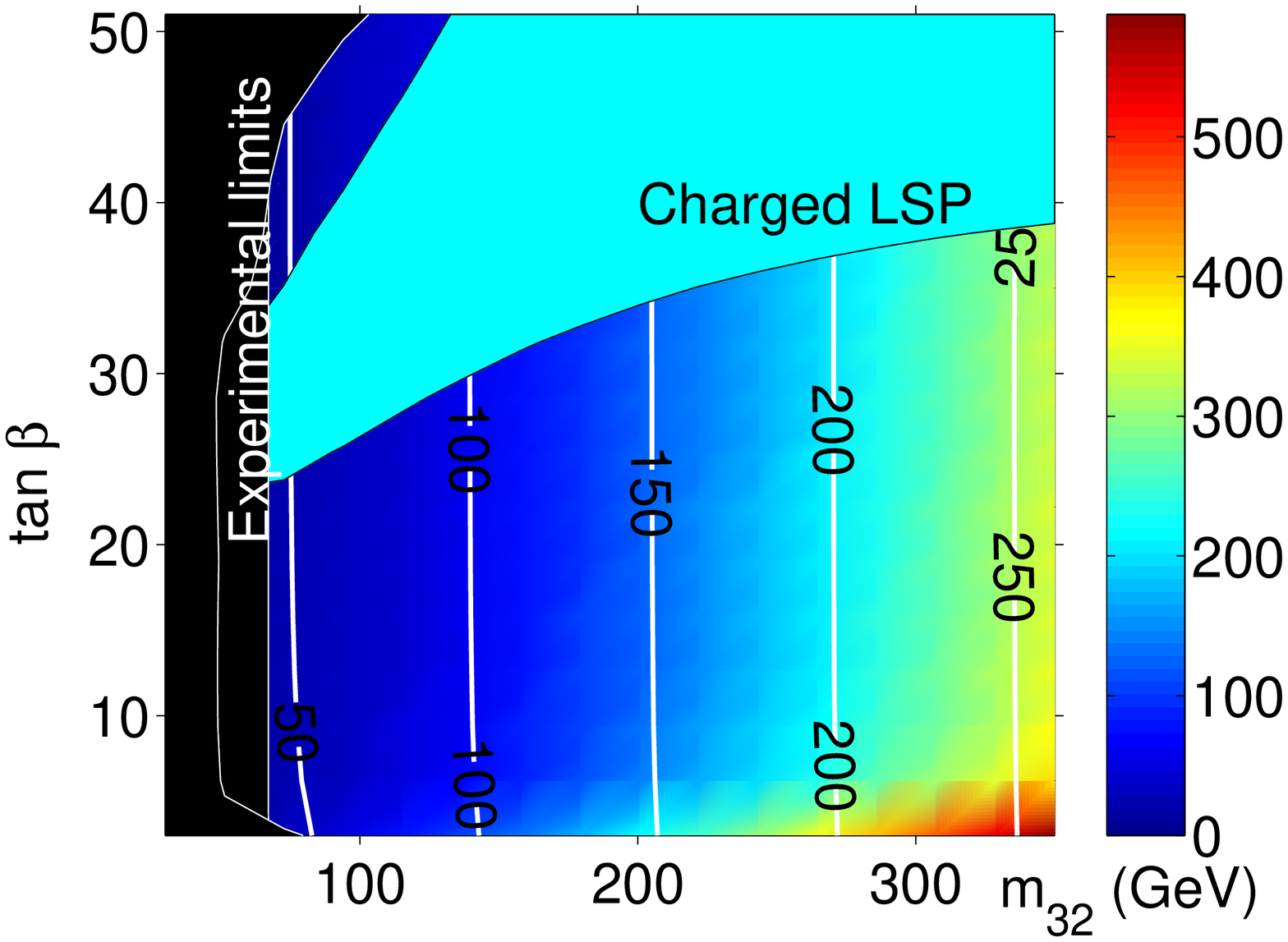}{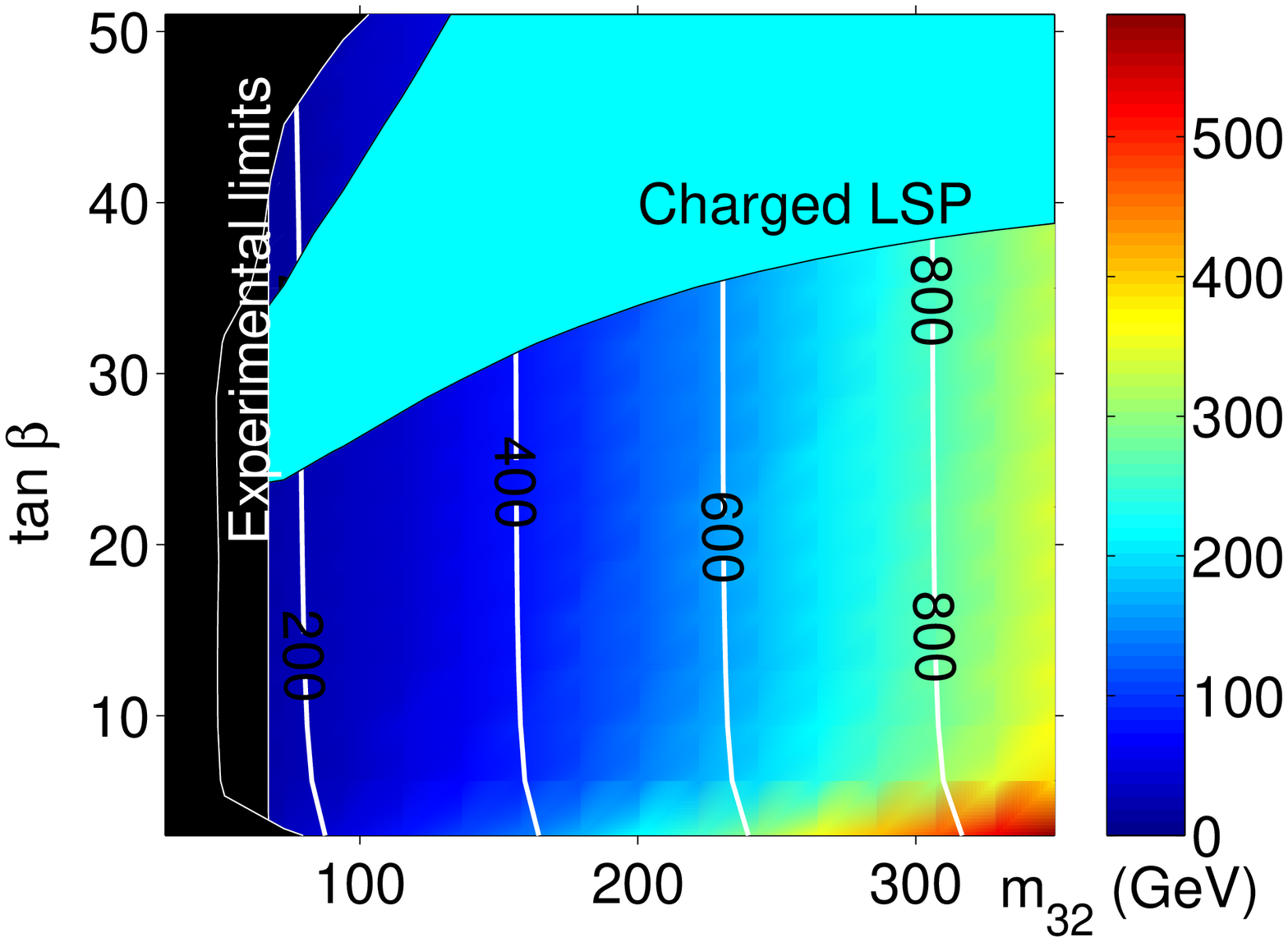}
\label{fig:gutdildom}
\caption{Spectra and fine tuning of GUT scale dilaton dominated
   scenario. $M_X=2
   \times 10^{16}$ GeV and sgn$(\mu)=+$. Fine tuning is
   displayed by the bar to the right. Note that the whole plane is ruled out
   by CCB constraints, but these have not been displayed. Regions of flat
   shading are ruled out by the labelled constraint.
   Contours of spectra (in 
   GeV) are shown for (a) $m_h$, (b) $m_{\tilde g}$, (c) $m_{{\chi}_1^0}$, (d)
   $m_{\tilde t_1}$.}
}
Fig.~\ref{fig:gutdildom} displays the fine-tuning over the whole
   dilaton-dominated parameter space, assuming the usual GUT scale $M_X =
   M_{GUT} =2
   \times 10^{16}$ GeV (and sgn($\mu$)=1). This value of $M_X$ 
   approximates the true perturbative heterotic string-scale $M_H\sim 5 \times
   10^{17}$ GeV. We have neglected the renormalization between $M_H$ and
   $M_{GUT}$, which should not be a bad approximation because the running
   depends logarithmically upon the renormalization scale, and we have only
   neglected one order of magnitude compared with 14 between the GUT scale and
   the weak scale.
The whole parameter space
   is {\em ruled out}\/ by the CCB constraint, as explained in the previous
   section. In order to compare the rest of the parameter space with a lower
   value of $M_X$, we do not display this constraint in the figure, but
   instead show the 
   spectra of the lightest Higgs $h$, the gluino ${\tilde g}$, the lightest
   neutralino $m_{\chi_1^0}$ and the lightest stop $m_{\tilde t_1}$. The
   experimental limits derive from the gluino and lightest MSSM Higgs
   constraints in table~\ref{bounds}. The region denoted `charged LSP' has the
   stau as the LSP and is therefore ruled out (if R-parity is conserved)\footnote{In this region, $m_{\chi_1^0}$ is almost degenerate with $m_{\tilde
   \tau}$. Thus, higher order radiative corrections could potentially raise
   the stau mass above the lightest neutralino. We therefore counsel care in
   the interpretation of this bound.}

\FIGURE[t]{%
\fourgraphs{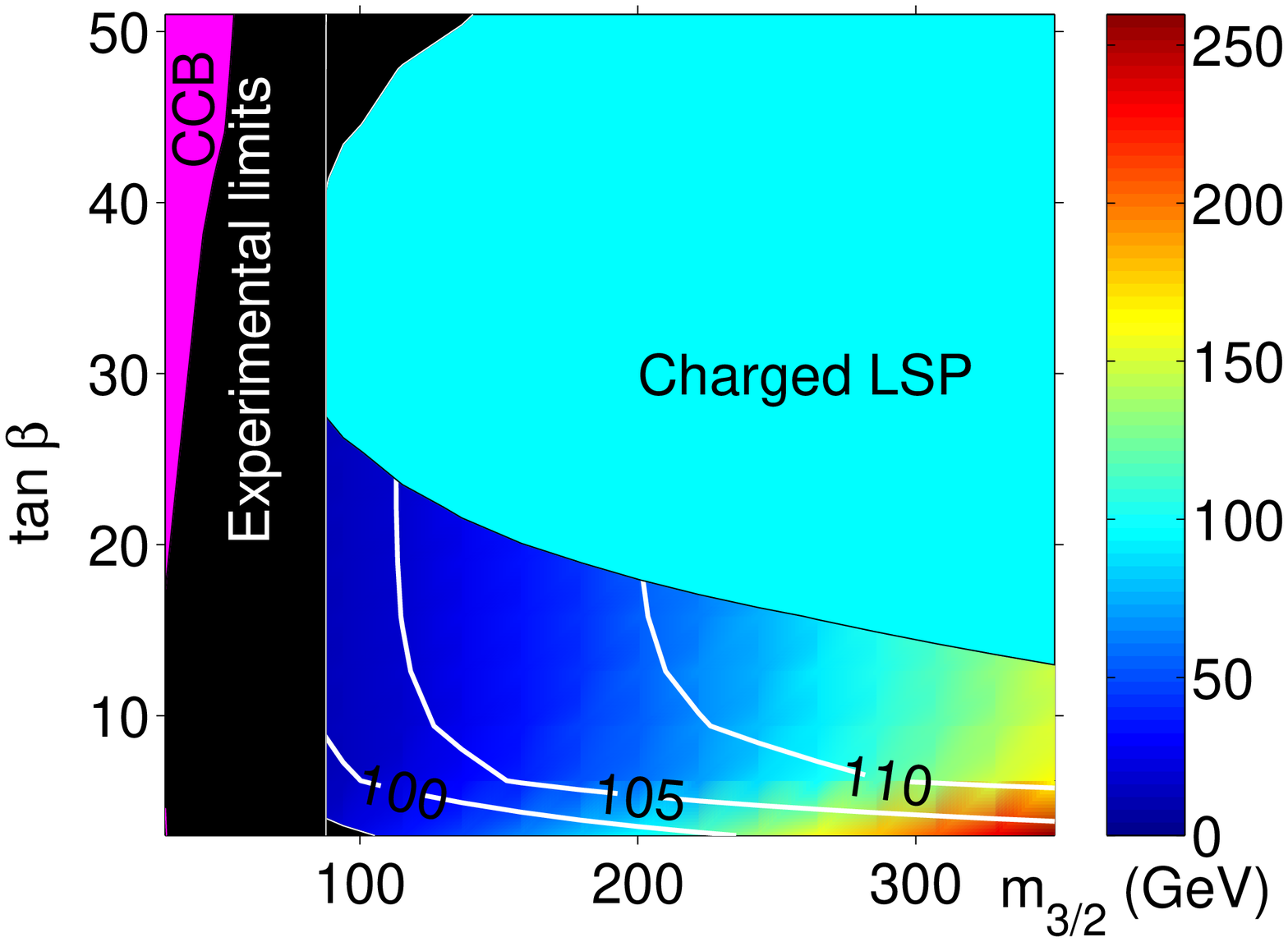}{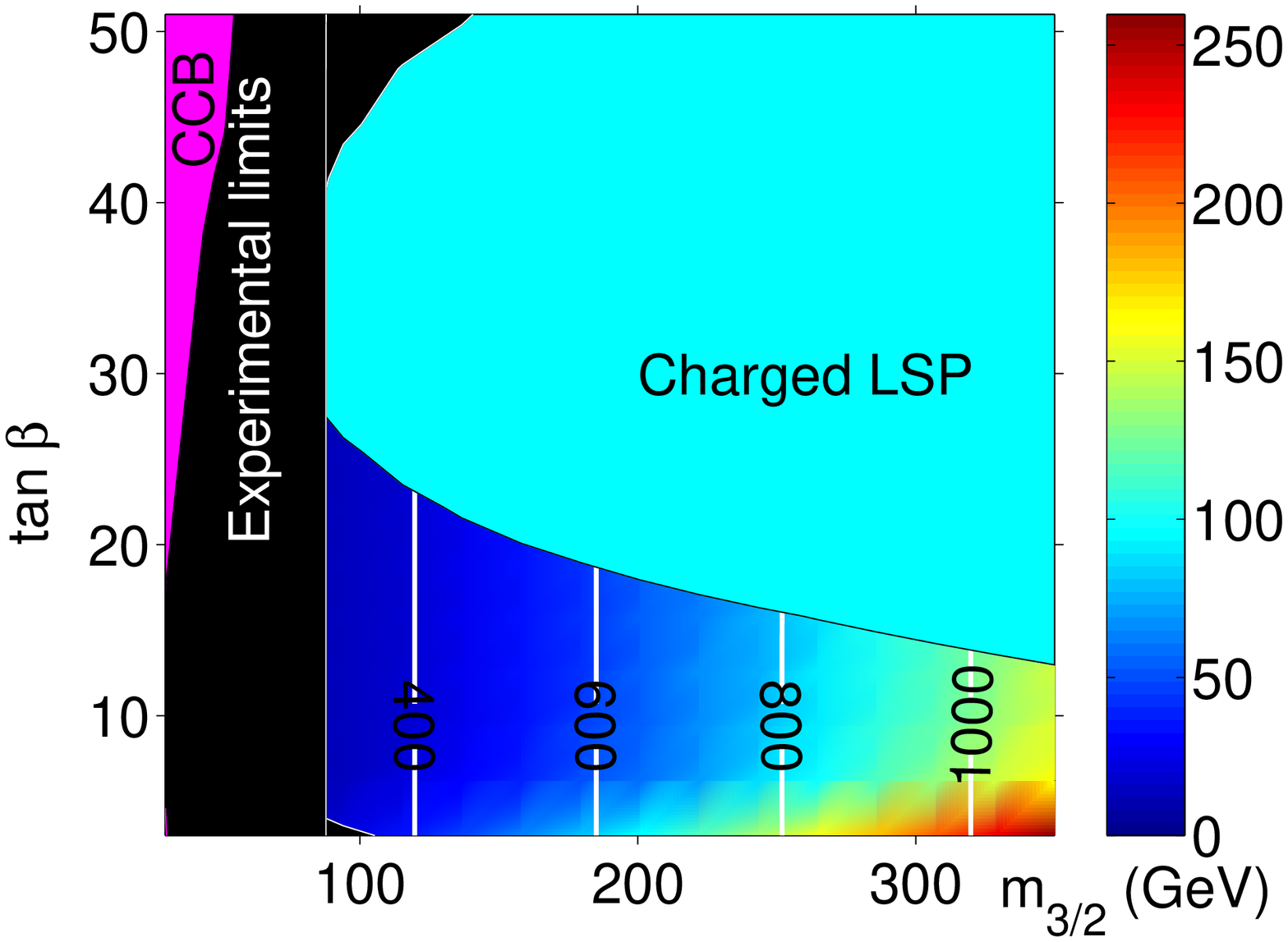}{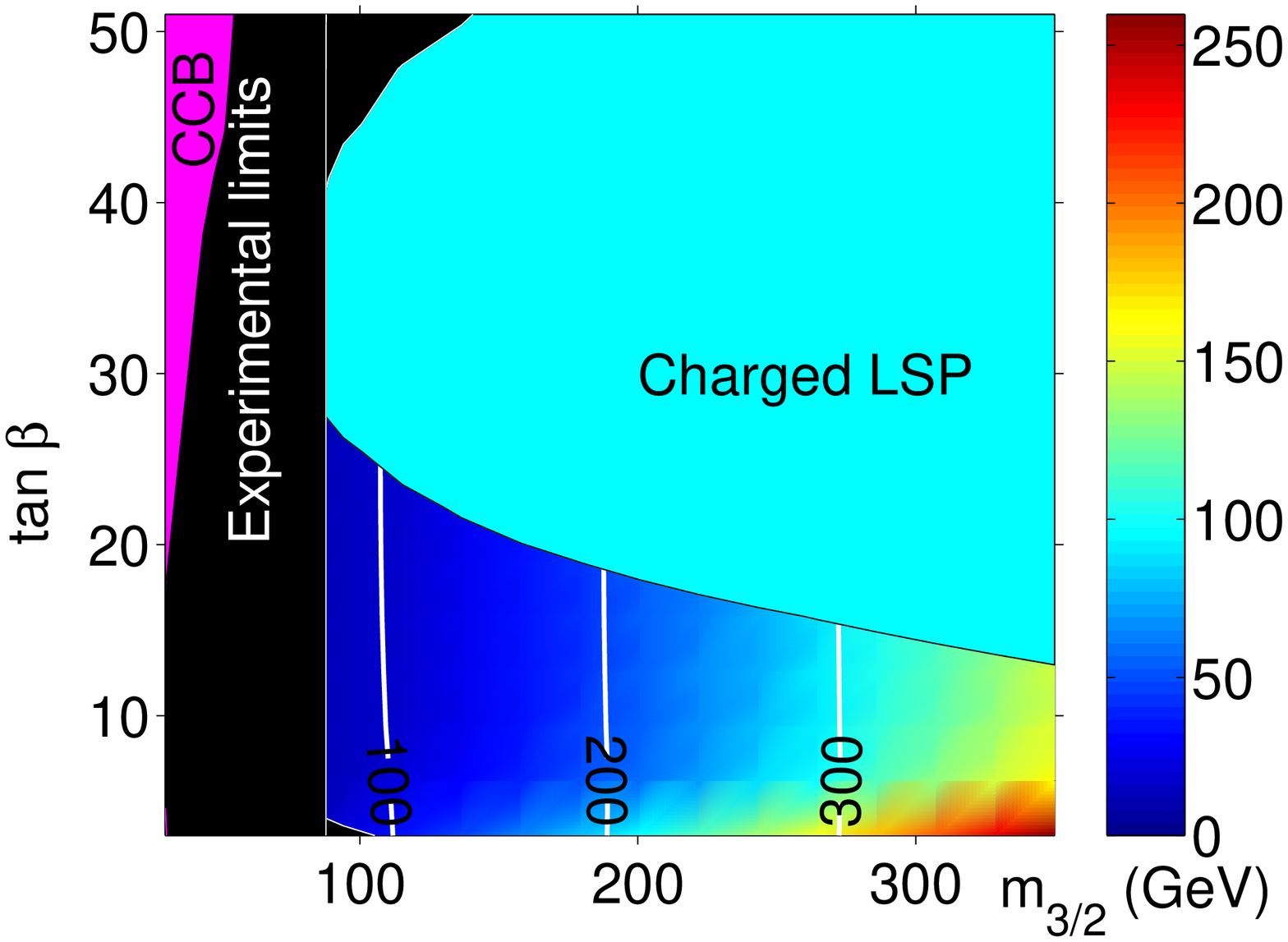}{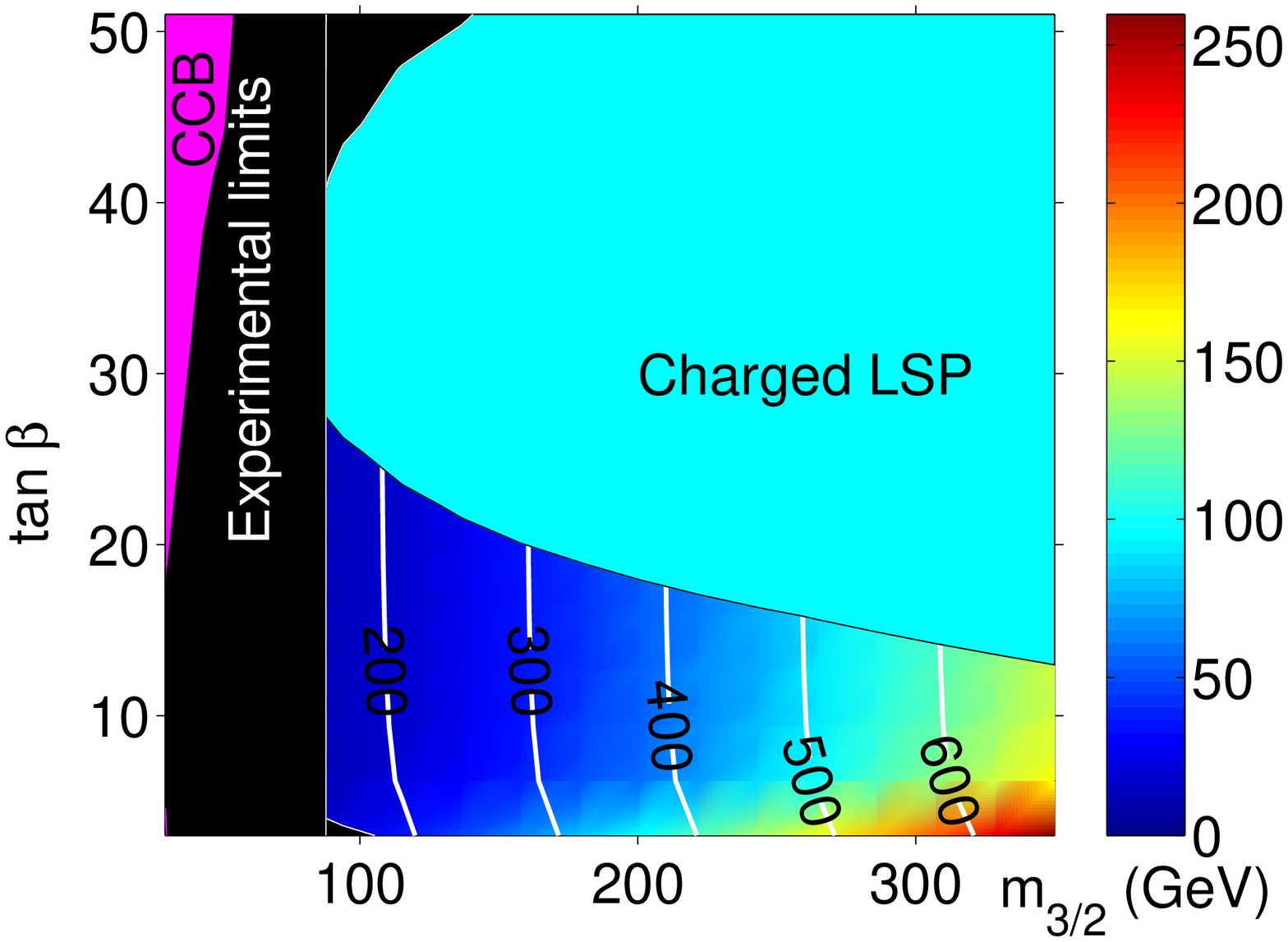}
\label{fig:intdildom}
\caption{Spectra and fine tuning of the intermediate scale dilaton dominated
   scenario. 
   $M_X=10^{11}$ GeV and sgn$(\mu)=1$. Fine tuning is
   displayed by the bar to the right. Regions of flat
   shading are ruled out by the labelled constraint.
   Contours of spectra (in 
   GeV) are shown for (a) $m_h$, (b) $m_{\tilde g}$, (c) $m_{{\chi}_1^0}$, (d)
   $m_{\tilde t_1}$.}
}
The fine-tuning parameter (as defined above) is displayed in the background
and by the bar to the right of each figure. The fine-tuning increases sharply
for the region of low $\tan \beta$ and high $M_{3/2}$. 
We have extended the region of parameter space for larger values of $M_{3/2}$
   than  those shown, and have found that $m_h<116$ GeV.

Fig.~\ref{fig:intdildom} shows the equivalent 
spectra and fine-tuning for an intermediate scale $M_X=10^{11}$ GeV and
sgn($\mu)$=1. 
It is important to note that the CCB bound, shown by the
shaded region, is now even less restrictive than the empirical lower bounds
upon the gluino and lightest MSSM Higgs.
We also see that $\tan \beta < 28$ is required by the charged LSP constraint.
	The
   fine-tuning is roughly half-that of the canonical GUT scale scenario, for a
   given point in parameter space. Extending the region of parameter space
   covered yields $m_h < 117$ GeV.

\FIGURE[t]{%
\label{fig:intdilsl}
\twographs{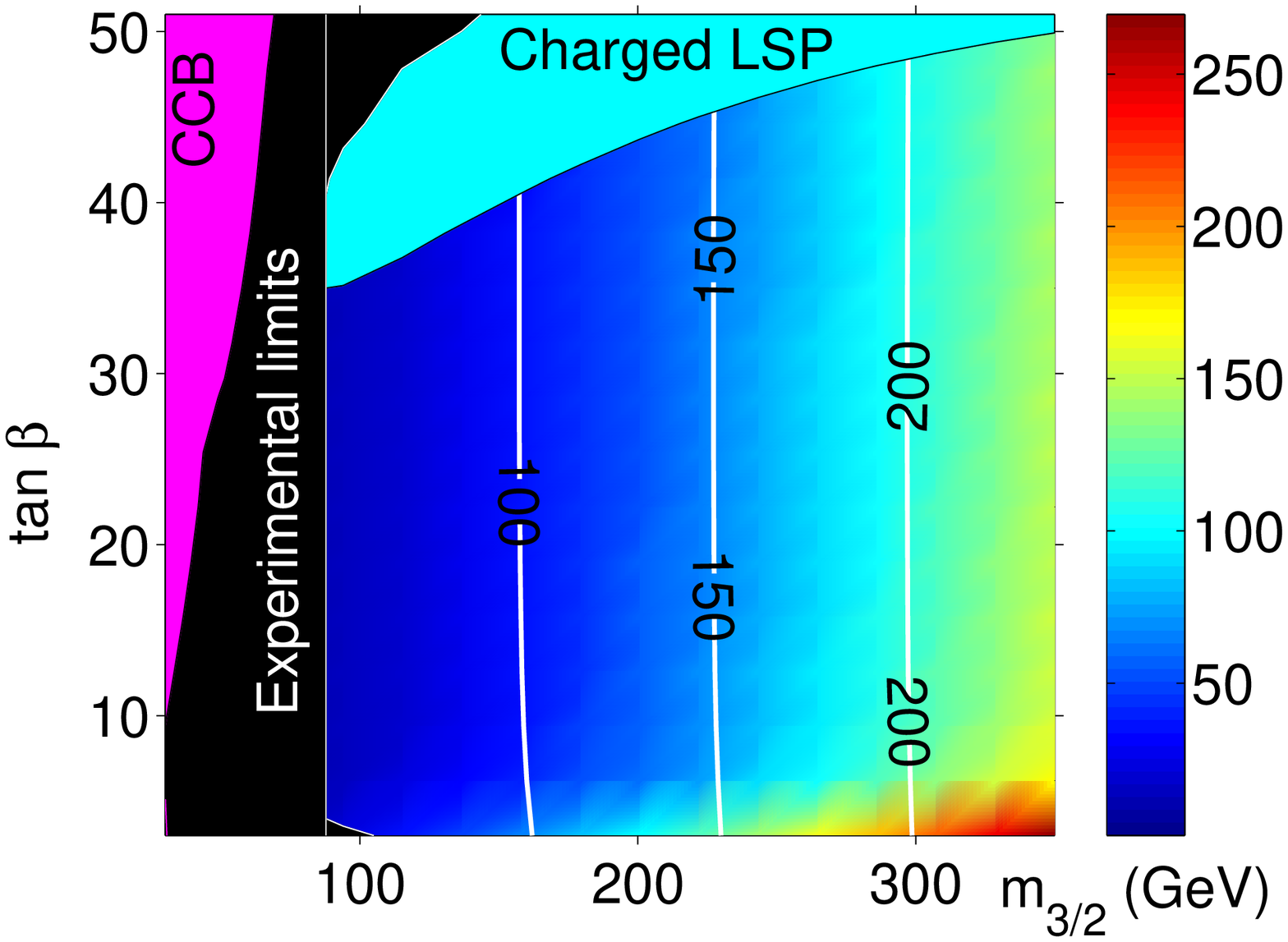}{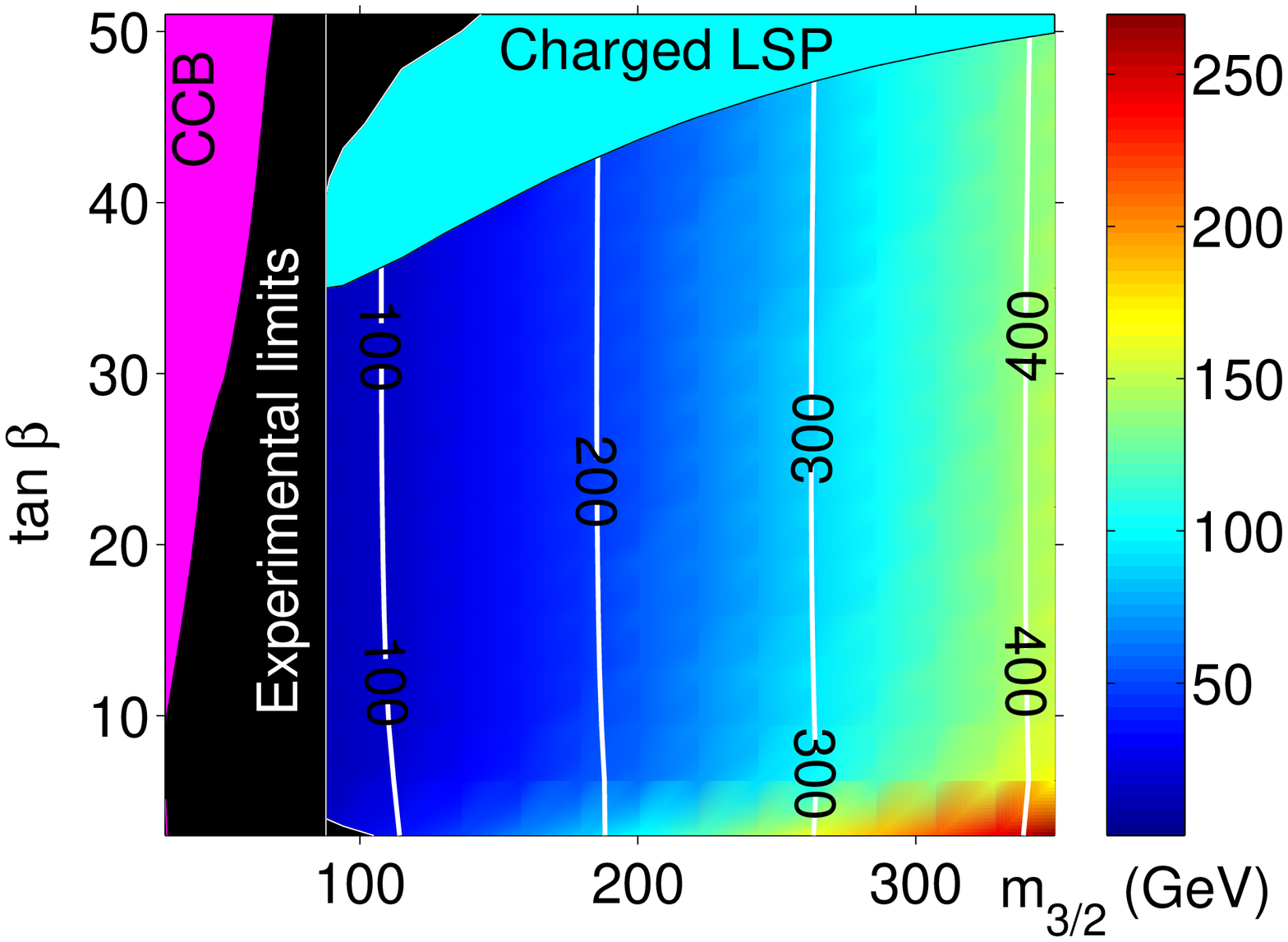}
\caption{Spectra and fine tuning of the {\em extra sleptons}\/ intermediate scale
   dilaton dominated 
   scenario, $M_X=10^{11}$ GeV and sgn$(\mu)=1$. The effects of $3 \times E_R
   + 2 \times L$ on the gauge couplings have been added at $m_t$ (to one-loop
   order) in order to display the possible effect of extra matter introduced
   to provide gauge unification.
	Fine tuning is displayed on the bar to the
   right. Flat shaded regions are excluded by the labelled constraints on the 
   figures. Contours of spectra    (in
   GeV) are shown for (a) $m_{{\chi}_1^\pm}$, (b) $m_{{\chi}_1^0}$. All other
   spectra are identical to the case without extra sleptons.}
}
We now examine the effect of adding extra states in order to achieve gauge
   unification at the intermediate scale. As a case study, we pick the example
   of extra leptons: $3 \times E_R + 2 \times L$ (and vector-like
   partners). We assume these extra states have negligible Yukawa couplings so
   that their effect upon the spectra may be encapsulated by changes in the
   beta functions of the gauge couplings only. We take these changes into
   account to one-loop order. There is negligible difference to any of the
   spectra except for the weak gauginos, and so we display the lightest
   chargino and neutralino masses in fig.~\ref{fig:intdilsl}. We note that, due
   to small corrections to the stau and lightest neutralino masses, the
   charged LSP bound has significantly relaxed.

\subsection{FCNCs and CP}

Flavour Changing Neutral Current processes (FCNCs) such as
$b\rightarrow s \gamma $ and Electric Dipole Moments
(EDMs) are important experimental tests for supersymmetry. 
Since a generic supersymmetric model fails these tests,
they offer important insight into the
structure of supersymmetry breaking. 
The dilaton dominated models in the 
present paper represent a significant improvement 
in this direction, as we now discuss.

First let us review the current status. 
It is generally believed that the observed absence of 
FCNCs and large EDMs implies that one or more of the following is 
an integral feature of the supersymmetry breaking
(see ref.\cite{vives} for a review); 

\begin{itemize}

\item Universality: This proposal uses the fact that the masses
of squarks and quarks can be simultaneously diagonalised 
if the supersymmetry breaking is degenerate for particles with the same 
hypercharge. The resulting suppression 
of FCNC is similar to the GIM mechanism.

\item Heavy 1st and 2nd generations~\cite{dim}: really a variant of the
first proposal which relies on the fact
that the most severe constraints tend to involve the 1st and 2nd generations,
whereas electroweak symmetry breaking involves the 3rd
generation. It is therefore  possible to make the 
squarks of the 1st and 2nd generation simultaneously 
heavy without paying too high 
a price in fine tuning. These models can, for example,  
be motivated by horizontal flavour symmetries.

\item Specific flavour structure~\cite{af}: a relaxation of the
first proposal which relies on the fact that 
EDMs and FCNCs depend on certain elements in
the supersymmetry breaking. For example,
EDMs can be acceptably small if the CP violation occurs only in flavour
off-diagonal elements of the supersymmetry breaking. This
type of situation can arise in heterotic string models with 
a suitable choice of modular weights. 

\end{itemize}

Concerning FCNC, 
the first of the above suggestions is the oldest and arises in the simpler
4 dimensional $N=1$ supergravity models\footnote{Note that as well as
these proposals there is of course the possibility that contributions to 
various processes happen to cancel~\cite{nath}.
This in general will involve 
 a certain amount 
of fine tuning.
}.
 In phenomenological
studies 
the model with degenerate $A$-terms and mass squareds (the Constrained
MSSM) has becomes something of a benchmark. However it is important to realise that 
such degeneracy only arises naturally in string theory in the 
dilaton dominated scenario
and as we have seen this scenario is excluded 
for theories in which the soft parameters are set at the 
conventional GUT scale because the physical vacuum is unstable. 
The remaining two proposals can therefore be seen as attempts 
to find a natural (in the sense of t'Hooft) solution to the 
SUSY flavour and CP problems that is consistent with
cosmological constraints.

However in the previous sections we have seen that one of 
the predictions of intermediate scale string scenarios 
is that the physical vacuum is stable in 
the dilaton dominated scenario and 
that fine-tuning is relatively mild. They are therefore 
the first realistic (string derived) models that 
automatically solve the SUSY 
flavour 
 problem, and are 
consistent with cosmological constraints. 

It is easy to see that in models with a lowered string scale 
the supersymmetric contributions to these 
FCNC processes are qualitatively
the same as in the CMSSM\@.
Running the RGEs in models with universal supersymmetry breaking 
at the string scale results in squark mass squareds 
that have small flavour off-diagonal components, $\Delta_{ij}$.
The phenomenological constraints are conveniently expressed as bounds on
\be 
\delta = \frac{\Delta}{\tilde{m}^2},
\ee
where ${\tilde{m}^2}$ is a measure of the average squark mass 
in the offending diagram.
One may use the mass insertion approximation to 
calculate bounds on these parameters (see \eg ref.\cite{gabbiani}).
For example $b\rightarrow s \gamma$ implies that 
$\left|(\delta _{23}^d)_{LR}\right| \leqsim 10^{-2}$ in order to satisfy the 
experimental bounds. Now, a typical contribution
to the off-diagonal piece is almost linear, \ie of the 
form $\Delta \sim \frac{A_{D23}}{M_3}$ (this of course is 
necessary for the mass insertion approximation to be valid at all). 
Renormalization 
contributions to $A_{D23}$ are proportional to $\log (M_X/M_W)$ and 
this implies that the FCNC effects are relatively independent 
of the string scale and qualitatively the same as those for 
standard SUGRA models. The quantitative differences will 
be discussed in detail in a future work~\cite{future}.

For CP violation, the dilaton domination scenario does not solve the
EDM problem, even though it has been argued that it can ameliorate it.
In the dilaton domination scenario, almost all the CP violation 
that we observe in the Kaon system, for instance, must be a result of 
CP violation in the Yukawa couplings. 
One remaining question for the dilaton
breaking scenario is how this CP violation arises. 
In string theory CP is a 
discrete gauge symmetry and consequently must be broken spontaneously.
A natural assumption in the dilaton dominated models is that this 
breaking is caused by the VEVs of moduli fields, even though 
they do not enter the supersymmetry breaking. 

This idea has been
examined for orbifold models in ref.\cite{bailin} and we 
briefly re-cap how CP violation appears. 
Ref.\cite{bailin}\  makes use of the  PSL(2,Z), $T$-duality
present in heterotic string models. In that case the scalar potential
for
the $T$ field always has supersymmetric 
 extrema at the points $T=1$ (real) and
$T=e^{i\pi/6}$ which breaks CP\@. If the minimum of the potential
is at any of these  points
$T$ will not break supersymmetry  and we have dilaton dominance. 
Therefore  if the minimum is at the CP violating fixed point
 $T=e^{i\pi/6}$,
this  will not induce CP violation on soft supersymmetry breaking terms,
 however 
CP violation enters the Yukawa couplings in a non-trivial way.

As we have said, for heterotic strings, 
the drawback is that 
dilaton dominance implies the existence of vacuum instability
and CCBs. However for intermediate scale string models this scenario 
of CP violation becomes extremely attractive. In particular it 
automatically excludes CP violation from all the supersymmetry breaking 
terms. 
There is an interesting
additional difference between the heterotic case and the present one.
In the models we have been discussing there is {\em no modular symmetry}. 
Thus the non-perturbative contributions to the 
superpotential coming from gaugino condensation are not restricted by 
modular invariance and consequently dilaton dominance 
plus spontaneous 
CP violation do not require us to be at the special fixed 
point $T=e^{i\pi/6}$. In the models being considered here, 
the only requirement is dilaton dominance. The expectation 
value of the modulus may be at any point consistent with 
this requirement and {\em generically}\/ this point will have a 
phase that spontaneously breaks CP. 

The reason why the dilaton domination scenario, through a mechanism
like this, does not really solve
the EDM problem is that it does not say anything about the phase of
the 
$\mu$ parameter (see for instance \cite{khalil}) which can then
contribute to CP violation. Therefore we can say that 
reviving the dilaton domination scenario, by having an intermediate
string scale, has the spin off of solving the FCNC problem and
may allow the possibility of ameliorating the EDM problem but does not
solve it.

\section{Conclusions}

\TABULAR[t]{cc|ccc}
{%
\label{tab:high}
SUSY Model & $M_X$/GeV & $c_{min}$ & $m_h^{max}$/GeV & 
CCB \\ \hline
MSSM& $2 \times 10^{16}$ GeV & 17.4 & 116  
& ${\mathbf \times}$ \\
MSSM & $10^{11}$ GeV & 8.8 &117  
& $\surd$ \\
MSSM plus leptons & $10^{11}$ GeV & 8.6 & 116 
& $\surd$ \\
}
{%
Highlights of phenomenological results in the various models.
Shown are: (a) $c_{min}$, the minimal fine-tuning parameter allowed by the
experiment and theoretical constraints, (b) the maximum lightest CP even Higgs
mass possible, (c) whether the global
minimum of the scalar potential is CCB (denoted by $\times$).
`Plus leptons' denotes the case where the MSSM spectrum, plus $3 \times E_R +
2 L_L$ vector-like representations is used.
}
Lowering the value of the fundamental scale clearly has very important
consequences for supersymmetric models. It makes available a new degree of
freedom that radically changes the nature and analysis of 
the low-energy implications of particular models through 
the running of the different physical
parameters under the renormalization group.
Our work is only the beginning of this exploration.
The most striking implication of our results is that the dilaton
domination scenario which was ruled out by the CCB constraint for high-scale
strings becomes viable when the string scale is lowered to the intermediate
scale. 
We summarise some highlights from the phenomenological results in
table~\ref{tab:high}. 
Dilaton
domination has reasonable fine-tuning, independent of whether or not
leptons are added in order to achieve field-theoretic gauge 
unification at the intermediate scale.

Here we have performed the analysis for an intermediate fundamental scale and
compared the results with the standard GUT scale scenario. 
These two particular scales are motivated by very different  
physical pictures. However, it may also be
interesting to repeat our analysis for fundamental scales 
in other ranges, even if they are not physically motivated at
present. An interesting possibility might be to start the running at 
a scale significantly closer to the TeV scale that, although leaving
some room for the running of the parameters, may still be relevant at
lower energies.

We have not explored the phenomenological prospects of 
all possible scenarios discussed in the text. In particular the new
possibility allowed in type I string models of having the 
blowing-up modes be the dominant source of supersymmetry breaking
may be worth exploring in the future, as well as other different
combinations.

Furthermore, recently a new class of D-brane models has been
constructed in which supersymmetry is explicitly broken and
transmitted to the observable sector via gravitational interactions,
 for which the intermediate scale is naturally selected
\cite{aiq, aiqu}. 
The most interesting models
in this class happen to be versions of the left-right symmetric
models, with $SU(3)\times SU(2)_L\times SU(2)_R$ gauge symmetry 
surviving at very low energies. They have spectra that give 
unification at the intermediate scale 
and other interesting properties,
such as automatic $R$-parity symmetry and a stable proton. It would be
very interesting to extend our analysis to include such
models.

In conclusion we believe that our analysis opens up new
avenues of exploration for supersymmetric models. Our
predictions allow the possibility of direct 
experimental verification in the near future.

\acknowledgments
This work has been partially supported by
 CICYT (Spain), the European Commission (grant
ERBFMRX-CT96-0045) and PPARC\@. 
Cambridge University High Performance Computing Facility has been used for the
numerical analysis. We would like to thank E. Poppitz and G.~Servant 
for conversations and K. Matchev
 for useful comparisons of
numerical computation.



\begin{thebibliography}{99}

\bibitem{witten} 
E.~Witten, \npb{471}{135}{1996}.

\bibitem{lykken}
J.~D.~Lykken, \prd{54}{1996}{3693}, hep-th/9603133. 

\bibitem{add}
N.~Arkani-Hamed, S.~Dimopoulos, G.~Dvali
\plb{429}{263}{1998}; \prd{59}{086004}{1998}.

\bibitem{aadd}
I.~Antoniadis, N.~Arkani-Hamed, S.~Dimopoulos, G.~Dvali
\plb{436}{257}{1998}.




\bibitem{benakli}
K.~Benakli,  \prd{60}{104002}{1999}, hep-ph/9809582.

\bibitem{biq}
C.~P.~Burgess, L.~E.~Ib\'a\~nez and F.~Quevedo,
\plb{447}{257}{1999},
hep-ph/9810535.


\bibitem{aiqu}
G.~Aldazabal, L.~E.~Ib\'a\~nez,  F.~Quevedo and A. Uranga,
hep-th/0005067.




\bibitem{aiq} 
G.~Aldazabal, L.~E.~Ib\'a\~nez and F.~Quevedo, 
hep-th/9909172 JHEP 0001 (2000) 031  [hep-ph/0001083].

\bibitem{keith}
K.~Dienes, E.~Dudas, T.~Gherghetta, \plb{436}{55}{1998},
hep-ph/9803466;
D. Ghilencea and G. Ross \plb{442}{165}{1998} [hep-ph/9809127];
\npb{569}{39}{2000} [hep-ph/9908369]; hep-ph/0001143.

\bibitem{mirage}
L.~E.~Ib\'a\~nez, hep-ph/9905349;
N.~Arkani-Hamed, S.~Dimopoulos and J.~March-Russell,   hep-th/9908146.

\bibitem{kim}
J.E. Kim, \prep{C150}{1987}{1}; astro-ph/9802061.

\bibitem{astroaxion}
M.~S.~Turner, \prep{C197}{1990}{67}; G.~Raffelt, \prep{C198}{1990}{1}.

\bibitem{axionX}
K.R. Dienes, E. Dudas and T. Gherghetta, \prd{62}{2000}{105023}, \hepph{9912455};

\bibitem{neutX}
K.R. Dienes, E. Dudas and T. Gherghetta, 
\npb{557}{1999}{25}, \hepph{9811428};
N. Arkani-Hamed {\em et al}, talk at {\em SUSY 98} conference, \hepph{9811448}.

\bibitem{lindekaloper}
N.~Kaloper and A.~Linde, 
\prd{59}{101303}{1999}, hep-th/9811141.
\bibitem{wimpzillas} 
E.~Kolb, D.~Chung and A.~Riotto, hep-ph/9810361.

\bibitem{orient}
A.~Sagnotti, in Cargese 87, {\it Strings  on Orbifolds}, ed. G.
Mack et al. (Pergamon Press, 1988) p. 521;
P.~Horava, \npb{327} {1989} {461}; \plb{231} {1989} {251};
J.~Dai, R.~Leigh and J.~Polchinski, Mod.Phys.Lett. A4 (1989) 2073; 
R.~Leigh, Mod.Phys.Lett. A4 (1989) 2767;
 G.~Pradisi and A.~Sagnotti, \plb{216}{1989}{59};
M.~Bianchi and A.~Sagnotti, \plb{247}{1990}{517};
E.~Gimon and J.~Polchinski, \prd{54}{1996}{1667}, hep-th/9601038.
E.~Gimon and C.~Johnson, \npb{477}{1996}{715}, hep-th/9604129;
A.~Dabholkar and J.~Park, \npb{477}{1996}{701}, hep-th/9604178.

\bibitem{imr}
L.~E.~Ib\'a\~nez, C.~Mu\~noz and S.~Rigolin, 
\npb{553}{43}{1999}, hep-ph/9812397.

\bibitem{ads} 
I.~Antoniadis, E.~Dudas and A.~Sagnotti,
\plb{464}{38}{1999};
G.~Aldazabal and A.~Uranga, JHEP 10 (1999) 024.

\bibitem{iru}
L.~E.~Ib\'a\~nez, R.~Rabadan and A.~Uranga, hep-th/9905098.

\bibitem{abd}
I.~Antoniadis, C.~Bachas and E.~Dudas,  \npb{560}{1999}{93}.

\bibitem{afiv} 
G.~Aldazabal, A.~Font, L.~E.~Ib\'a\~nez and G.~Violero,
\npb{536}{1998}{29}.

\bibitem{poppitz} 
E.~Poppitz, \npb{542}{1999}{31}.

\bibitem{iq} 
L.~E.~Ib\'a\~nez and F.~Quevedo, JHEP 10 (1999) 001,
 hep-ph/9908305.

\bibitem{bim}
A.~Brignole, 
L.~E.~Ib\'a\~nez and C.~Mu\~noz,
\npb{442}{1994}{125}.  

\bibitem{benakli2}
K. Benakli, hep-ph/9911517.

\bibitem{amsb}
L.~Randall and R.~Sundrum, hep-th/980155; G.~Giudice, M.~Luty,
H.~Murayama and
R.~Rattazzi, JHEP 9812 (1998) 027;
T.~Gherghetta, G.~F.~Giudice and J.~D.~Wells, \npb{559}{1999}{27}
\hepph{9904378}, J. Bagger, T. Moroi and E. Poppitz, hep-ph/9911029.


\bibitem{stringpheno}
B.~de~Carlos, J.~A.~Casas and C.~Mun\~oz, \plb{299}{1993}{234};
Ph.~Brax, U.~Ellwanger and C.~A.~Savoy, \plb{347}{1995}{269};
S.~Khalil, A.~Masiero and F.~Vissani, \plb{375}{1996}{154};
A.~Love, P.~Stadler, \npb{515}{1998}{34};
G.~Cleaver {\em et al}, \prd{59}{1999}{115003};
D.~G.~Cerdeno and C.~Mun\~oz, \prd{61}{2000}{016001};
T.~Kobayashi, J.~Kubo, H.~Shimabukuro, \hepph{9904201};
H.~Baer, M.~A.~Diaz, P.~Quintana, X.~Tata, \hepph{0002245}.

\bibitem{sakisf}
A.~Dedes and A.~E.~Faraggi, \hepph{9907331}.

\bibitem{susyguts}
U.Amaldi et al., \prd{36}{1987}{1385};
P.Langacker and M.Luo, \prd{44}{1991}{817};
J.Ellis,S.Kelley and D.V.Nanopoulos, \plb{249}{1990}{441}; \npb{373}{1992}{55};
U.Amaldi, W. de Boer and H.Fustenau, \plb{260}{1991}
447;
C. Giunti, C.W. Kim and U.W. Lee, \mpla{6}{1991}{1745};
H.Arason et al., \prd{46}{1992}{3945};
F.Anselmo, L.Cifarelli, A.Peterman and A.Zichichi, {\em Nuovo Cimento} {\bf
105 A}
(1992) 1179;
P.Langacker and N.Polonski, \prd{47}{1993}{4028};
A.E.Faraggi and B.Grinstein, \npb{422}{1994}{3}.

\bibitem{MandV}
S. Martin and M. Vaughn, \prd{50}{1994}{2282} \hepph{9311340}.

\bibitem{yukun}
M. Chanowitz, J. Ellis and M. Gaillard, \npb{128}{1977}{506}; A. Buras,
J. Ellis, M. Gaillard and D.V. Nanopoulos, 
\npb{293}{1978}{66}; M.B. Einhorn and D.R.T. Jones, \npb{196}{1982}{475};
J. Ellis, D.V. nanopoulos and S. Rudaz, \npb{202}{1982}{43}; J. Ellis,
S. Kelley, and D.V. Nanopoulos, \npb{373}{1992}{55}; 
V. Barger, M.S. Berger and P. Ohmann, \prd{47}{1993}{1093};
M. Carena, S. Pokorski and C.E.M. Wagner, \npb{406}{1993}{59};
N. Polonsky and P. Langacker, \prd{50}{1994}{2199};

\bibitem{largecor}
L.J. Hall, R. Rattazzi and U. Sarid, \prd{50}{1994}{7048};
M. Carena {\em et al}, \npb{419}{1994}{213}, 
\npb{426}{1994}{269}.



\bibitem{ccb1}
J-M.~Fr\`ere, D.~R.~T.~Jones and S.~Raby, \npb{222}{11}{1983};
M.~Claudson, L.~Hall and I.~Hinchcliffe, \npb{228}{501}{1983};
H-P.~Nilles, M.~Srednicki and D.~Wyler, 
\plb{120}{346}{1983};
J-P.~Derendinger and C.~A.~Savoy, \npb{237}{307}{1984};
H.~Komatsu, \plb{215}{323}{1988};
P.~Langacker and N.~Polonsky, \prd{50}{2199}{1994};
A.~Kusenko, P.~Langacker and G.~Segr\'e, \prd{54}{5824}{1996};
J.~A.~Casas and S.~Dimopoulos, \plb{387}{107}{1996};
J.~A.~Casas, \hepph{9707475}

\bibitem{casas1}
J.~A.~Casas, A.~Lleyda and C.~Mu\~noz, \npb{471}{3}{1996}

\bibitem{dilaton}
J.~A.~Casas, A.~Lleyda and C.~Mu\~noz, \plb{380}{59}{1996}

\bibitem{ccb2}
J.~A.~Casas, A.~Lleyda and C.~Mu\~noz, \plb{389}{305}{1996};
H.~Baer, M.~Brhlik, D.~Castano, \prd{54}{6944}{1996};
T.~Falk, K.~A.~Olive, L.~Roszkowski and 
M.~Srednicki, \plb{367}{183}{1996};
A.~Riotto and E.~Roulet, \plb{377}{60}{1996};
A.~Strumia, \npb{482}{24}{1996};
T.~Falk, K.~A.~Olive, L.~Roszkowski, A.~Singh and 
M.~Srednicki, \plb{396}{50}{1997}; 
S.~A.~Abel and B.~C.~Allanach,
\plb{431}{339}{1998}; 
S.~A.~Abel and C.~A.~Savoy, \npb{532}{3}{1998};
S.~A.~Abel and T.~Falk, \plb{444}{427}{1998}

\bibitem{as98}
S.~A.~Abel and C.~A.~Savoy, \plb{444}{119}{1998}

\bibitem{casas2}
J.~A.~Casas, A.~Ibarra and C.~Mu\~noz, \npb{554}{67}{1999}

\bibitem{ccb3}
I.~Dasgupta, R.~Rademacher and P.~Suranyi,
\plb{447}{284}{1999};
U.~Ellwanger and C.~Hugonie,
\plb{457}{299}{1999}

\bibitem{sandb}
S.~A.~Abel and B.~C.~Allanach,
\hepph{9909448}


\bibitem{measures}
See for example
R. Barbieri and G. Giudice, \npb{306}{1988}{63};
 R. Barbieri and A. Strumia, \plb{433}{1998}{63};
B. de Carlos and J.A. Casas, \plb{320}{1993}{320}.

\bibitem{focuspoints}
J.L Feng, K.T. Matchev and T. Moroi, \hepph{9909334}; \hepph{9908309}

\bibitem{rs}
A. Romanino and A. Strumia, \hepph{9912301}

\bibitem{BBO}
V. Barger, M.S. Berger and P. Ohmann, \prd{49}{4908}{1994}.

\bibitem{sakisandlah}
A. Dedes, A.B. Lahanas and K. Tamvakis, \prd{53}{1996}{3793}.

\bibitem{zwirneretal}
G. Gamberini, G. Ridolfi and F. Zwirner, \npb{331}{1990}{331}.

\bibitem{georgetc}
S. Heinemeyer, W. Hollik and G. Weiglein, \plb{455}{1999}{179},
\hepph{9903404}.

\bibitem{PDG}
Particle Data Group, {\em Eur. Phys. Jnl.} {\bf C3} (1998) 1.


\bibitem{vives}
A.~Masiero, O.~Vives, 
{\it Int. School Subnuclear Physics}, Erice, Italy, 29 Aug - 7 Sep 1999,
[hep-ph/0003133] 

\bibitem{dim}
S.~Dimopoulos and G.~F.~Giudice, \plb{357}{573}{1995};
A.~Cohen, D.~B.~Kaplan and A.~E.~Nelson, \plb{388}{599}{1996};
A.~Pomarol and D.~Tommasini, \npb{466}{3}{1996}

\bibitem{af}
S.~A.~Abel and J.~M.~Frere,
Phys.\ Rev.\  {\bf D55} (1997) 1623
[hep-ph/9608251];
S.~Khalil, T.~Kobayashi  and A.~Masiero, 
Phys.\ Rev.\  {\bf D60} (1999) 075003 [hep-ph/9903544];
S.~Khalil, T.~Kobayashi, 
Phys.\ Lett.\  {\bf B460} (1999) 341 [hep-ph/9906374]



\bibitem{nath}
M.~Brhlik, L.~Everett, G.~L.~Kane and
J.~Lykken, hep-ph/9908326;
E.~Accomando, R.~Arnowitt and B.~Dutta,
 Phys.\ Rev.\ {\bf D61}, 075010 (2000)  [hep-ph/9909333];  
M.~Brhlik, L.~Everett, G.~L.~Kane, S.~F.~King and O.~Lebedev, 
Phys.\ Rev.\ Lett.\ {\bf 84}, 3041 (2000) 
[hep-ph/9909480];
T.~Ibrahim and P.~Nath,
Phys.\ Rev.\ {\bf D61}, 093004 (2000) 
[hep-ph/9910553]

\bibitem{gabbiani}
F.~Gabbiani, E.~Gabrielli, A.~Masiero and L.~Silvestrini,
Nucl.\ Phys.\  {\bf B477} (1996) 321
[hep-ph/9604387].

\bibitem{future}
Work  in progress

\bibitem{bailin}
D.~Bailin, G.~V.~Kraniotis and A.~Love,
\plb{432}{1998}{90}; \plb{435}{1998}{323};
\plb{463}{1999}{174}



\bibitem{khalil}
S.M. Barr and S. Khalil, \prd{61}{2000}{035005} [hep-ph/9903425].








\end{thebibliography}
\end{document}